\documentclass[traditabstract]{aa}
\usepackage{graphicx,psfig}
\newcommand{\Tstar} {T$_{\rm{eff}}$}

\newcommand{\degree} {$^{\rm o}$}

\newcommand{\simless}{\mathbin{\lower 3pt\hbox
      {$\rlap{\raise 5pt\hbox{$\char'074$}}\mathchar"7218$}}} 
\newcommand{\simgreat}{\mathbin{\lower 3pt\hbox
     {$\rlap{\raise 5pt\hbox{$\char'076$}}\mathchar"7218$}}} 

\begin{document}

\newcommand{\hi}{\ion{H}{i}~}
\newcommand{\hii}{\ion{H}{ii}~}

   \title{Dust Properties of Protoplanetary Disks in the Taurus-Auriga \\ Star Forming Region from Millimeter Wavelengths}

   \author{L. Ricci         \inst{1}
\and          L. Testi          \inst{1,2} \and A. Natta \inst{2}
\and R. Neri \inst{3} \and S. Cabrit \inst{4} \and G. J. Herczeg \inst{5}}


   \institute{  European Southern Observatory,
   Karl-Schwarzschild-Strasse 2, D-85748 Garching, Germany
            \and
                     INAF - Osservatorio Astrofisico di Arcetri, Largo Fermi 5, I-50125 Firenze, Italy
            \and
            Institut de Radioastronomie Millim\'etrique, 300 Rue de la Piscine, F-38406 Saint Martin d'H\`eres, France
            \and LERMA, UMR 8112 du CNRS, Observatoire de Paris, 61 Av. de l'Observatoire, F-75014 Paris, France
            \and Max Planck Institut fur Extraterrestrische Physik, Postfach 1312, D-85741 Garching, Germany}

   \date{Received 2009 October 5/ Accepted 2009 December 15}

   \titlerunning{Protoplanetary Disks in Taurus-Auriga}

   \authorrunning{Ricci et al.}

\abstract{
We present the most sensitive 3 mm-survey to date of protoplanetary disks carried in the Taurus-Auriga star forming region (average rms of about 0.3 mJy), using the IRAM PdBI. With our high detection rate of 17/19, we provide the first detections at wavelengths longer than about 1 mm for 12 sources. This enables us to study statistically the mm SED slopes and dust properties of faint disks and compare them to brighter disks using a uniform analysis method.
With these new data and literature measurements at sub-millimeter and millimeter wavelengths, we analyze the dust properties of a sample of 21 isolated
disks around T Tauri stars in the Taurus-Auriga star forming region.
Together with the information about the disks spatial extension from sub/mm-mm interferometric studies, we derive from the observed sub-mm/mm spectral energy distribution constraints on the dust opacity law at these wavelengths, using two-layer flared disk models and a self-consistent dust model that takes properly into account the variation of the dust opacity with grain growth. We find evidence for the presence in the disk midplane of dust particles that have grown to sizes as large as at least 1 millimeter in all the disks of our sample, confirming what was previously observed on smaller brighter objects. This indicates that the dust coagulation from ISM dust to mm-sized grains is a very fast process in protoplanetary disks, that appears to occur before a young stellar object enters the Class II evolutionary stage. Also, the amount of these large grains in the disk outer regions is stationary throughout all the Class II evolutionary stage, indicating that mechanisms slowing down the dust inward migration are playing an important role in the Taurus-Auriga protoplanetary disks. Another result is that the spectral index between 1 and 3 mm for the 6 faintest disks in our sample is on average smaller than for the brighter disks, indicating either that these fainter, yet unmapped, disks are spatially much less extended than the brighter spatially resolved disks, or that fainter disks have typically larger dust grains in their outer regions. Considering that these fainter disks are more representative of the bulk of the disk population than the brighter ones, this may have important consequences for the theories of planetesimal formation and disk formation and evolution. Finally, we investigate the relations between the derived dust properties, namely dust mass and grain growth, and the properties of the central star, like its mass, age and mass accretion rate.
}



\maketitle


\section{Introduction}

Circumstellar disks play a fundamental role in the physical processes involved in star and planet formation. In these systems part of the material (gas and dust) loses angular momentum and accretes onto the central forming star, while another part of it may give birth to a planetary system. Although the content of dust is only a small fraction of the overall material in a circumstellar disk, dust grains are crucial elements in the early stages of planet formation. According to the core-accretion scenario (Safronov \& Zvjagina~\cite{Saf69}, Pollack et al.~\cite{Pol96}), growth from an initial ISM population of submicrometer-sized dust grains to planetesimals of $1-100$ km sizes is believed to be the key mechanism of early planet formation. These planetesimals ``embryos'' then continue to grow leading to the formation of at least terrestrial planets and possibly the cores of giant planets.

Observational evidence of grain growth from the ISM dust has been obtained from a variety of techniques at different wavelengths. Throop et al.~(\cite{Thr01}) observed in optical and near-infrared the translucent edge of the 114-426 protoplanetary disk in the Orion Nebula Cluster, and derived from the extinction curve of the background nebular emission a lower limit of a few microns for the dust grains in the very outer regions of the disk, about 500 AU away from the central star.
Another method to probe the presence of large grains in the disk is through the shape and intensity of the 10- and 20-$\mu$m silicate features. The data indicate a large variety of silicate profiles, ranging from strongly peaked silicate bands and steeply rising spectral energy distributions (SEDs) to ``boxy'' silicate profiles and flatter SEDs (ISO, Malfait et al.~\cite{Mal98}; Spitzer Space Telescope, Kessler-Silacci et al.~\cite{Kes06}). The boxy features with low feature-to-continuum ratios are interpreted as grain growth to micron size (Bouwman et al.~\cite{Bou01}).

However, the techniques outlined so far can only probe dust grains as large as some $\mu$m in the disk surface layers.  They are not sensitive to larger grains in the disk midplane, the region in which planet formation is supposed to occur. In order to probe larger grains in the midplane, observations at millimeter wavelengths are needed.
Beckwith \& Sargent~(\cite{Bec91}) showed that T Tauri stars have shallow SEDs at sub-millimeter wavelengths. Under the assumption of optically thin emission at these frequencies, this implies a dust opacity dependence on wavelength ($\kappa_{\rm{mm}} \propto \lambda^{-\beta}$) much flatter than in the ISM ($\beta_{\rm{ISM}} = 1.7$), which is naturally interpreted in terms of grain growth (see e.g. Draine~\cite{Dra06}). However, this interpretation of the disk SEDs is not unique, since the same data can be explained by very small optically thick disks. To break this degeneracy and sort out the effect of potentially large optical depth it is necessary to spatially resolve the disks to determine their actual sizes. Furthermore, in this context observations at millimeter wavelengths are very useful, since at these lower frequencies the impact of optically thick disk inner regions to the total emission is expected to be lower. Therefore, to actually probe the dust properties in protoplanetary disks, one needs to combine the determination of the sub-millimeter/millimeter SED with information on the disk extension from high-angular resolution interferometric observations (see e.g. Testi et al.~\cite{Tes01}).

In the last years, several sub-mm/mm interferometric observations have been carried out to investigate dust grain growth in protoplanetary disks. Wilner et al.~(\cite{Wil00}) resolved the disk around the TW Hya pre-main-sequence (PMS) star at 7 mm using the Very Large Array (VLA). Extensive modelling of the SED of this source has shown that the dust grains in the outer parts of the TW Hya disk have grown to at least $\sim 1$ cm (Calvet et al.~\cite{Cal02}). An analogous result has been obtained for the disk around CQ Tau (Testi et al.~\cite{Tes03}). More recently, Rodmann et al.~(\cite{Rod06}) resolved 10 disks in the Taurus-Auriga star forming region and found clear evidence of grain growth in 4 of them. Lommen et al.~(\cite{Lom07}) resolved 1 disk in Chamaeleon and 4 in Lupus with the Submillimeter Array (SMA) at 1.4 mm and with the Australia Telescope Compact Array (ATCA) at 3.3 mm, and found clear evidence of dust grain growth to sizes of a few millimeter for 4 of them. Finally Schaefer et al.~(\cite{Sch09}) observed a sample of 23 low mass PMS stars (with spectral types of K7 and later) in Taurus-Auriga at 1.3 mm and 2.6 mm with the Plateau de Bure Interferometer (PdBI); they detected only 8 sources at 1.3 mm and 6 at 2.6 mm, and found evidence of grain growth for the 3 disks that they could spatially resolve.

In this paper we present our analysis on 21 protoplanetary
disks around T Tauri stars in the Taurus-Auriga
star forming region without stellar companions in the range $0.05''-3.5''$. For 11 of these objects, mainly faint disks with $F_{\rm{1.3mm}} < 100$ mJy, we have obtained new data
at $\sim$ 3 mm with the Plateau de Bure Interferometer\footnote{The Plateau de Bure
Interferometer at IRAM is supported by INSU/CNRS (France), MPG
(Germany) and IGN (Spain).} (hereafter PdBI) with an average rms
of about 0.3 mJy. Together with the data already present in the literature at sub/mm- and mm- wavelengths, and with the information obtained in the last years on the disks spatial extension from high-angular resolution interferometric observations at mm wavelengths, we investigate the dust properties in the disks, namely grain growth and dust mass, and their relation with the properties of the central star.

In Section \ref{sec:obs_sample} we present our new PdBI data, the properties of our sample, the method we used to estimate the main stellar physical quantities, and the data we used for our analysis. In Section \ref{sec:models} we describe the disk models adopted for the analysis of the disks sub-mm/mm SED. In Section \ref{sec:results} we show and discuss the results of our study in terms of dust grain growth and dust mass, whereas in Section \ref{sec:summary} we summarize our main findings.

\section{New 3mm observations and sample properties}
\label{sec:obs_sample}

\subsection{New PdBI observations}
\label{sec:obs}

We observed a total list of 19 targets; 16 of them were selected to be relatively faint, with $15$ mJy$< F_{\rm{1.3mm}} < 100$ mJy and with no known detections beyond about 1mm; the other 3 (HL, DO and DR Tau) are brighter sources that were observed for references purposes. The observations were carried out with PdBI between
July and August, 2007. Owing to antenna maintenance work, all observations
were carried out in a subarray of the six-element interferometer. The
antennas were arranged in very compact configurations that provided
sensitive baselines from 15\,m to 80\,m. The dual-polarization receivers,
which were observing in the 3mm band, had typical receiver temperatures of
about 35\,K and were providing image sideband rejection values better than
10\,dB. To achieve maximum sensitivity in the continuum, the spectral
correlator was adjusted to cover a total effective bandwidth of about
2\,GHz.

The observing procedures were set up to fill in gaps in the scheduling
without any specific tuning wavelength within the 3mm band. Visibilities
were obtained using on-source integration times of 22\,min interspersed with
2\,min calibrations on a nearby calibrator. Each star was observed on one or
more occasions for a minimum of 40\,min and a maximum of 2\,hr on-source.
The atmospheric phase stability on the baselines was always better than
40$^\circ$, consistent with seeing conditions of $1''$-$3''$, typical for
summer. The absolute flux density scale was calibrated on 3C84,
0528+134, MWC349 and Mars, and was found to be accurate to 10\%. The
receiver passband shape was determined with excellent precision on a strong
calibrator like 3C84.

We used the GILDAS package for the data reduction and analysis. The data
were calibrated in the antenna based manner. The continuum visibilities were
gridded with natural weighting and no taper to maximize the sensitivity.
Since none of the target stars was found to be resolved in synthesized beams
of $2''$-$4''$, a point source model was fitted to the calibrated
visibilities of each star to estimate the flux density of the continuum in
the 3mm band. The results are summarized in Table \ref{tab:obs}. Among the 19 targets, 17 were detected, including 14 of the 16 faint sources, thanks to our good sensitivity; 12 of these are first detections longward of about 1mm, strongly expanding the available database of 3mm data towards fainter disks. The 2.6mm detections of LkHa 358 and GO Tau were recently published by Schaefer et al.~(\cite{Sch09}).

In Sections \ref{sec:sample}, \ref{sec:stars}
and \ref{sec:mdot} we describe the final sample used for our analysis of dust properties, the adopted method to estimate the
stellar properties and mass accretion rates summarized in Table
\ref{tab:sample}, whereas in \ref{sec:mm-data} we report the sub-mm/mm
data that we collected from the literature for our analysis.

\begin{table*}
\centering \caption{ Summary of the new PdBI observations. } \vskip 0.1cm
\begin{tabular}{lrrrrrrr}
\hline \hline

\\
Object  & $\alpha$ (J2000)  & $\delta$ (J2000) &   $\nu$   & $\lambda$ & $F_{\nu}$ & rms   & Comments$^1$ \vspace{1mm} \\
        &                   &                  &   (GHz)   &  (mm)     &   (mJy)       & (mJy) &           \\
\\
\hline
\\
CW Tau  &  04:14:17.0     &   28:10:56.51 &  84.2  &  3.57 & 3.44  & 0.31 & \vspace{1mm} \\
CX Tau  &  04:14:47.8     &   26:48:11.16 &  86.2  &  3.49 & 1.01  & 0.13 & \vspace{1mm} \\
DE Tau  &  04:21:55.6     &   27:55:05.55 &  92.6  &  3.23 & 3.32  & 0.25 & \vspace{1mm} \\
DK Tau  &  04:30:44.3     &   26:01:23.96 &  84.2  &  3.57 & 4.70  & 0.20 & $2.5''$ binary \vspace{1mm} \\
DO Tau  &  04:38:28.6     &   26:10:49.76 &  84.2  &  3.57 & 13.80 & 0.29 & \vspace{1mm} \\
DP Tau  &  04:42:37.7     &   25:15:37.56 &  84.2  &  3.57 & $<$ 0.78 & 0.26 & no sub-mm info \vspace{1mm} \\
DR Tau  &  04:47:06.2     &   16:58:43.05 &  100.8 &  2.97 & 13.90 & 0.47 & \vspace{1mm} \\
DS Tau  &  04:47:48.6     &   29:25:10.96 &  100.8 &  2.97 & 2.96  & 0.21 & \vspace{1mm}  \\
FM Tau  &  04:14:13.5     &   28:12:48.84 &  100.8 &  2.97 & 2.65  & 0.24 & \vspace{1mm}  \\
FP Tau  &  04:14:47.4     &   26:46:26.65 &  100.8 &  2.97 & 11.80 & 0.26 & no sub-mm info \vspace{1mm} \\
FY Tau  &  04:32:30.5     &   24:19:56.69 &  100.8 &  2.97 & 1.90  & 0.28 & no sub-mm info \vspace{1mm} \\
FZ Tau  &  04:32:31.8     &   24:20:02.53 &  100.8 &  2.97 & 2.89  & 0.24 & \vspace{1mm}  \\
GK Tau  &  04:33:34.6     &   24:21:06.07 &  100.8 &  2.97 & $<$ 0.96 & 0.32 & $2.5''$ binary \vspace{1mm} \\
GO Tau  &  04:43:03.1     &   25:20:17.38 &  86.2  &  3.49 & 4.05  & 0.34 & \vspace{1mm}  \\
Haro 6-28 & 04:35:56.9    &   22:54:36.63 &  87.0  &  3.45 & 0.92  & 0.18 & $0.7''$ binary \vspace{1mm} \\
HL Tau  &  04:31:38.4     &   18:13:57.37 &  84.5  &  3.53 & 43.40 & 0.68 & flat spectrum \vspace{1mm} \\
HO Tau  &  04:35:20.2     &   22:32:13.98 &  87.0  &  3.45 & 2.03  & 0.53 & \vspace{1mm}  \\
LkHa 358 & 04:31:36.2     &   18:13:43.20 &  84.5  &  3.53 & 2.46  & 0.22 & no sub-mm info \vspace{1mm} \\
SU Aur  &  04:55:59.4     &   30:34:10.39 &  84.5  &  3.53 & 2.88  & 0.19 &
\vspace{2mm}
\\

\hline

\end{tabular}
\begin{flushleft}
1) Reason why the source has not been considered in the analysis (see text for more details).
\end{flushleft}
\label{tab:obs}

\end{table*}

\subsection{Sample}
\label{sec:sample}

Our final sample of 21 sources consists of all young stellar objects (YSO) in
Taurus-Auriga catalogued in Andrews \& Williams~(\cite{And05}) that fulfill the following
criteria: (1) a Class II infrared SED, in order to avoid contamination of sub-mm
fluxes by a residual envelope; (2) information on the central star through optical-NIR spectroscopic/photometric data, necessary to calculate self-consistent disk
SED models; (3) one detection at $\sim$ 3mm (either from our new PdBI observations
or from the literature) and at least one detection in the $0.45$ mm $ < \lambda < 0.85$ mm
spectral region. This selection criterion was chosen in order to probe
a broad enough spectral window to efficiently constrain the optically thin part of
the sub-mm/mm spectral energy distribution (SED); (4) no evidence of stellar
companions in the $0.05 - 3.5''$ range in angular separation. This range corresponds
to about $5 - 500$ AU in projected physical separation at the Taurus-Auriga star forming region distance, here assumed to be 140
pc for all the sources of our sample (Bertout et al.~\cite{Ber99}). The limit of 3.5 arcsec
ensures that mm interferometric flux measurements, of typical resolution 1
arcsec, are not contaminated by the companion's disk, while binaries closer
than 5 AU should have outer disk properties similar to those of single stars on
the scales probed by (sub-)mm data.

Only 21 sources catalogued by Andrews \& Williams~(\cite{And05}) fulfill the above 4
criteria. They are listed in Table \ref{tab:sample}, and include 11 sources from our PdBI
sample\footnote{The 8 sources dropped from the PdBI sample are HL Tau (a
flat spectrum source that retains an envelope; Padgett et al.~\cite{Pad99}); DP Tau, FP
Tau, FY Tau and LkHa 358 that have no available submm detections for SED
fitting; and DK Tau, GK Tau and Haro 6-28 which are binary systems with angular
separations of 2.5$''$ (Simon et al.~\cite{Sim92}), 2.5$''$
(Reipurth \& Zinnecker~\cite{Rei93}) and 0.7$''$ (Leinert et al.~\cite{Lei93}) respectively. We keep in
our sample the wide binaries DS Tau and HO Tau (angular separations of 7.1$''$
and 6.9$''$ respectively, Mathieu et al.~\cite{Mat94}); they do not
show any emission from the companions from our new
3 mm interferometric observations. The other dropped multiple systems from the
AW05 catalog are DH Tau (2.3$''$ binary, Itoh et al.~\cite{Ito05}) , GG Tau (quadruple
system with a circumbinary ring of inner and outer radii of 180 and 260 AU respectively,
Pinte et al.~\cite{Pin07}), and UY Aur (0.9$''$ binary, Leinert et al.~\cite{Lei93})}.

Figure 1 reports histograms that describe the completeness level of our final
sample with respect to all Taurus-Auriga Class II YSOs catalogued in Andrews \&
Williams~(\cite{And05}), and fulfilling the ``isolation'' criterion (4). Our sample contains
63\% of these ``isolated'' class II showing a flux at $\sim$ 0.85mm greater than $100$ mJy,
and 31\% of the sources with lower 0.85mm fluxes. Therefore, our sample is not
complete even for the brightest YSOs at 0.85mm. This is due to a lack of
observations at
$\sim$ 3mm for these sources. However, thanks to the high sensitivity of our new PdBI
observations (see Section \ref{sec:obs}) our sample enables us to study
statistically for the first time the dust properties of faint
disks (i.e. with $F_{\rm{1.3mm}} < 100$ mJy, $F_{\rm{0.85mm}} < 100$ mJy)
and compare them to the brighter ones using a uniform
analysis method.
In terms of the stellar properties,
it is important to note that the Andrews \& Williams~(\cite{And05}) catalogue includes only a
few very low mass PMS stars: only stars with an estimated
mass higher than about 0.2 $M_{\odot}$ have been detected because of
the sensitivity limits of the current facilities at sub-mm
and mm wavelengths that make the detection of disks
around very low mass YSOs very difficult. The sensitivity limitation is even more
severe at 3mm, therefore our final sample
includes 58\% of all the class II
isolated sample of Andrews \& Williams~(\cite{And05}) with an estimated mass greater than 0.4 $M_{\odot}$, but only
two sources with $M_{\star} < 0.4$ $M_{\odot}$  (out of 10 in Andrews \& Williams~\cite{And05}).

\begin{figure*}[htbp!]
 \centering
\begin{tabular}{cc}
\includegraphics[scale=0.45]{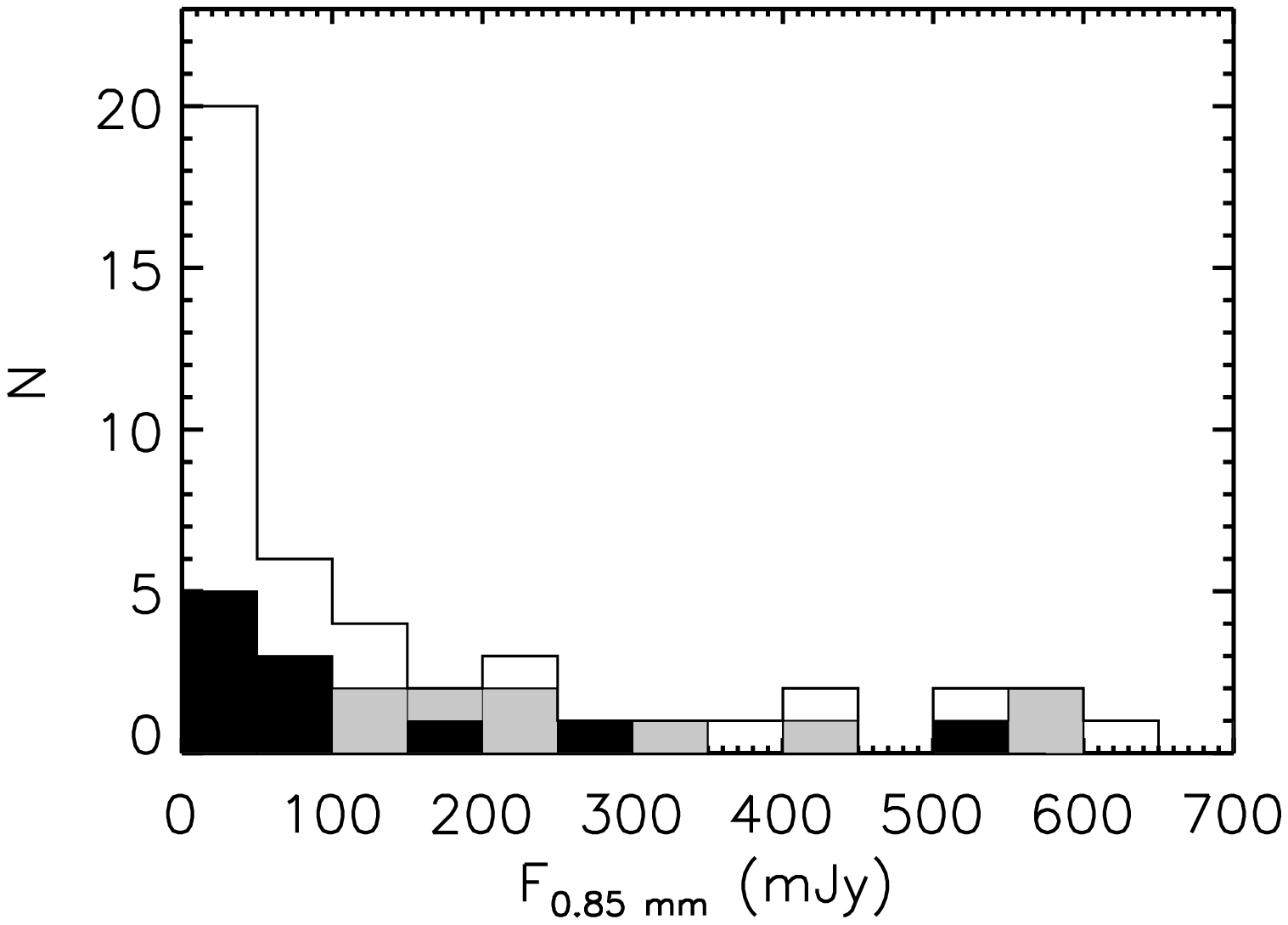} &
\includegraphics[scale=0.45]{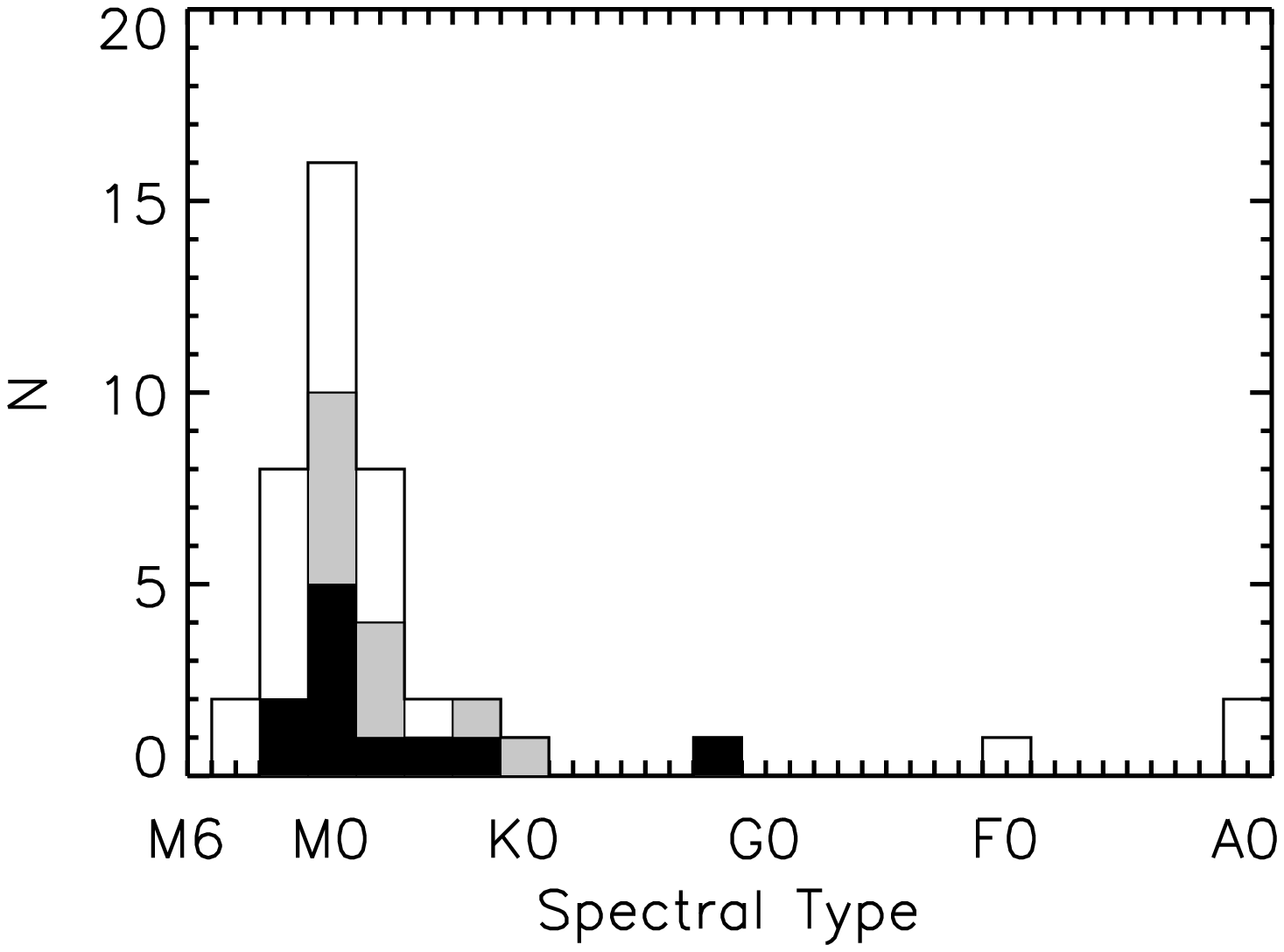} \\
 \multicolumn{2}{c}{\includegraphics[scale=0.45]{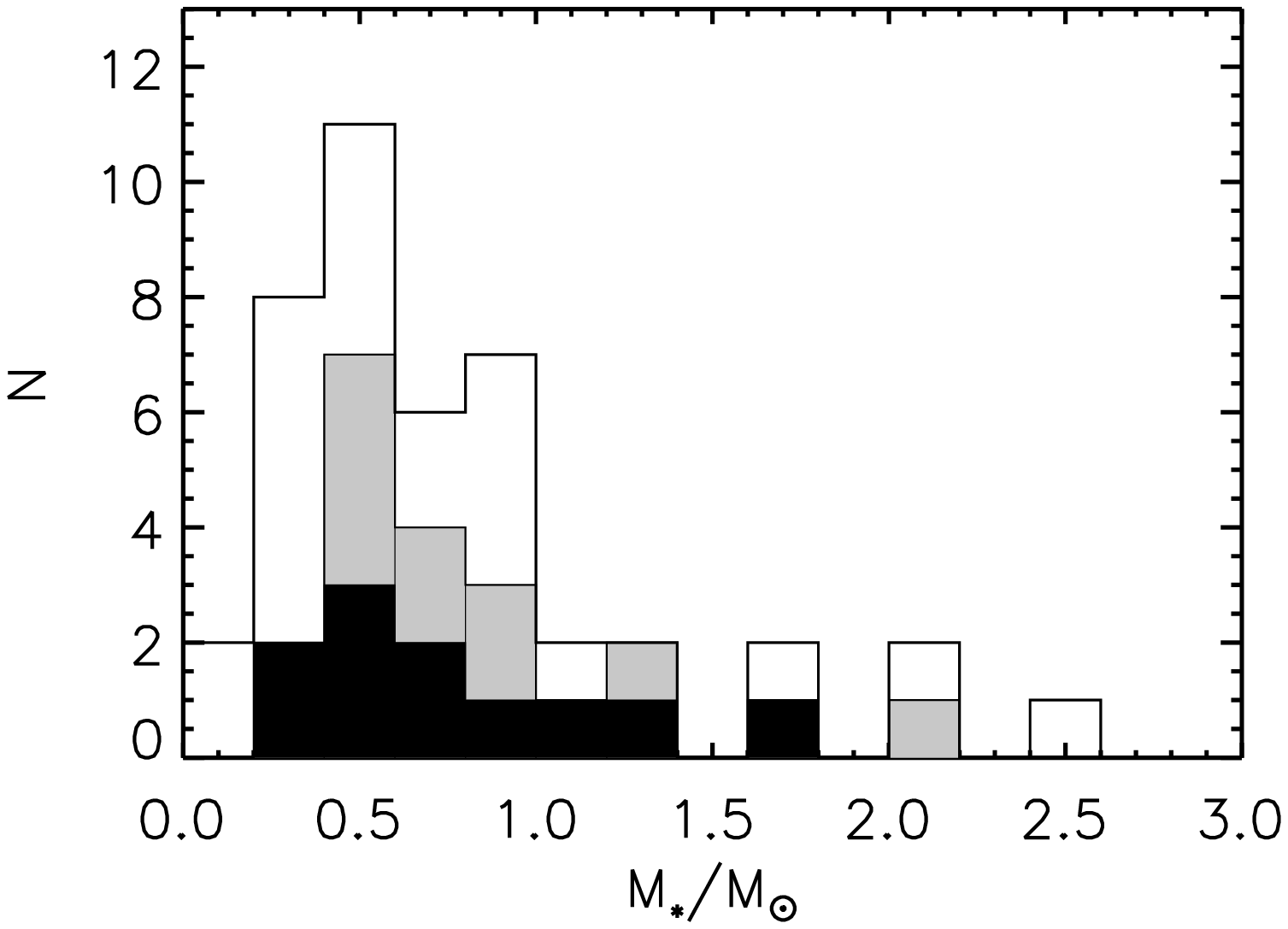}}

 \end{tabular}
 \caption{Histograms higlighting some properties of our selected sample. In all the histograms our sample is represented by black$+$grey columns, the black columns represent the disks in our sample for which we obtained new PdBI data at $\sim$ 3 mm, while the total columns (black$+$grey$+$white) include all the class II YSOs from the Andrews \& Williams~(\cite{And05}) catalogue with no evidence of stellar companions in the $0.05-3.5''$ interval in angular separation, from which our sample has been selected (see text). Upper left histogram represents the distribution of the $\sim$0.85 mm fluxes, including upper limits, for all the sources but two, 04113+2758 and GM Aur, that have not been observed at $\sim$0.85 mm. Upper right and bottom histograms show the distribution of stellar spectral type and estimated stellar mass (see Section \ref{sec:stars}) respectively. Four class II YSOs are not included in these histograms because for them an estimate of the stellar spectral type is not available. These are 04301+2608, FT Tau, Haro 6-39 and IC 2087.}

 \label{fig:completeness}
\end{figure*}

\subsection{Stellar properties}
\label{sec:stars}


Spectral types for all the sources in our sample are taken from the literature (Kenyon \& Hartmann~\cite{Ken95},
Brice\~no et al.~\cite{Bri98}) and cover a range from G2 (SU Aur) to M2.5 (CX Tau).
Spectral types of M0 and earlier have been converted to effective
temperatures with the dwarf temperature scale of Schmidt-Kaler~(\cite{Sch82}),
whereas for types later than M0 we adopted the temperature scale
developed by Luhman~\cite{Luh99}. Typical uncertainties for these sources
are $\sim 1-1.5$ in spectral sub-types or $\sim 100-150$ K in
temperature.

The stellar luminosities of all the pre-main-sequence (PMS) stars
were computed from their $J$-band flux (2MASS All Sky Catalog of point
sources, Cutri et al.~\cite{Cut03}) to minimize contamination from UV and IR
excess emission. For the bolometric corrections we adopted the
dwarf values from Kenyon \& Hartmann~(\cite{Ken95}) that are considered to be
satisfactory approximations for these young sources in the
$J$-band (Luhman~\cite{Luh99}).

In order to estimate the amount of extinction toward these young
sources, we calculated the color excesses using intrinsic colors
provided by Kenyon \& Hartmann~(\cite{Ken95}) for G2$-$K7 spectral types and by
Leggett~(\cite{Leg92}) for M0$-$M2.5 types. To ensure that the color excess
reflects only the effect of reddening, minimizing the emission due to accretion, the typical selected colors
are between the $R$ and $H$ bands. Since for some of the objects
at later spectral types in our sample $R-I$ measurements are
not available, we dereddened the $J-H$ and $H-K_{s}$
colors from the 2MASS All Sky Catalog of point
sources\footnote{For this analysis the $JHK_{s}$ magnitudes have been
transformed from the 2MASS photometric system to the Johnson-Glass
one using the color transformations reported in Carpenter~(\cite{Car01}) and
Bessell \& Brett~(\cite{Bes88}).} to the locus observed for classical T Tauri stars
(CTTS) by Meyer et al.~(\cite{Mey97}), following the method described in
Brice\~no et al.~(\cite{Bri02}). Extinctions are finally calculated adopting the
extinction law of Rieke \& Lebofsky~(\cite{Rie85}).

The computed photospheric luminosities cover a range from $0.16 \ L_{\odot}$
(HO Tau) to $10.6 \ L_{\odot}$ (SU Aur). Considering the uncertainties
in the photometry, reddenings, bolometric corrections and
distance, the typical errors in the bolometric luminosities are
$\pm 0.08-0.13$ in $\log L_{\star}$. Slight differences from the
values obtained by other authors are typically within the uncertainties and
are not significant for the purposes of this paper.

\begin{figure}
 \centering
 \resizebox{\hsize}{!}{\includegraphics{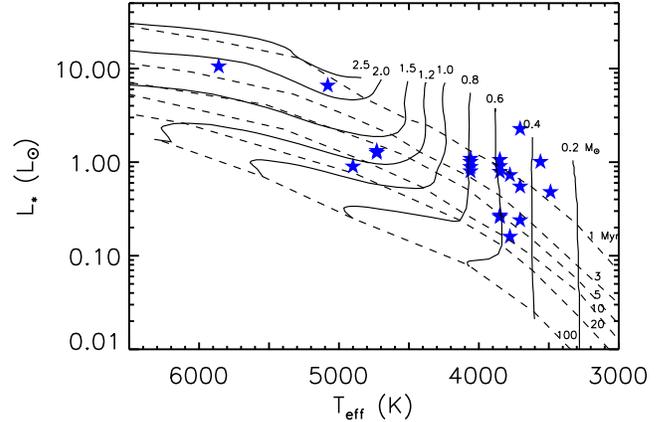}}
 \caption{H-R diagram for the sources of our sample. The dashed lines correspond to the isochrones for ages in Myr as labeled at the right end of the lines from the \cite{Pal99} PMS evolutionary models, while the solid ones represent the evolutionary tracks from the same models for PMS stars with masses as labeled at the top of the evolutionary track lines. In this diagram the evolutionary tracks start from an age of $0.1$ Myr. The errorbars for the ($T_{\rm{eff}}$, $L_{\star}$) values of our objects are not shown, the uncertainties in the photometric parameters are typically $\sim 0.08-0.13$ in $\log L_{\star}$ and $\sim 100-150$ K in temperature.}
 \label{fig:hrd}
\end{figure}

Given the effective temperatures and photospheric luminosities, we
placed the PMS stars on the H-R diagram and derived stellar masses
and ages adopting the theoretical tracks and isocrones of
Palla \& Stahler~(\cite{Pal99}) (see Figure \ref{fig:hrd}). One important source of
uncertainty for stellar masses and ages is given by the spread of
values obtainable by using different PMS evolutionary models. For
example, if compared to the ones obtained with the Baraffe et al.~(\cite{Bar98})
models, our adopted values of stellar masses and ages are
typically lower by a factor of $\sim 1.5$ and $\sim 2$
respectively. Our choice of using the Palla \& Stahler~(\cite{Pal99}) models is
uniquely due to the more complete coverage of the HR diagram
plane, that allows us to use the same evolutionary models for all the
sources of our sample. One should always bear in mind the high
uncertainty associated to these quantities, expecially to the
stellar age (see Section \ref{subsec:spectral_slopes} for a more deailed discussion about the estimates of YSO ages). According to the Palla \& Stahler~(\cite{Pal99})
models the ranges spanned by our sample go from about $0.3 \ M_{\odot}$
(CX Tau) to $2.2 \ M_{\odot}$ (RY Tau) in stellar masses and from
about $0.1$ Myr (UZ Tau E) to $17$ Myr (DS Tau) in stellar ages.

The main stellar quantities described here are summarized in Table
\ref{tab:sample}.

\begin{table*}

\centering \caption{ Stellar properties. } \vskip 0.1cm
\begin{tabular}{lcrrrcrrrr}
\hline \hline
\\
Object  & ST  & \Tstar & $L_{\star}$ & $L_{\rm{acc}}$ & Ref. ($L_{\rm{acc}}$)$^{\rm{a}}$ & $R_{\star}$ & $Log \ \dot{M}_{\rm{acc}}$ & $M_{\star}$ & Age \vspace{1mm} \\
&     &   (K)  &  ($L_{\odot}$)  & ($L_{\odot}$)  &                     &  ($R_{\odot}$)  & ($M_{\odot}$ yr$^{-1}$)  & ($M_{\odot}$) & (Myr)   \\
\\
\hline
\\
AA Tau  &   K7     &   4060  &  0.80 & 0.03 &  1    & 1.8 & -8.56  & 0.8 & 2.4           \vspace{1mm} \\
CI Tau  &   K7     &   4060  &  1.01 & 0.11 &  2    & 2.0 & -7.95  & 0.8 & 1.7           \vspace{1mm} \\
CW Tau  &   K3     &   4730  &  1.32 & 0.05 &  2    & 1.7 & -8.58  & 1.2 & 6.6        \vspace{1mm} \\
CX Tau  &   M2.5   &   3488  &  0.48 & 0.03 &  2    & 1.9 & -8.18  & 0.3 & 0.8           \vspace{1mm} \\
CY Tau  &   M1     &   3705  &  0.55 & 0.04 &  1    & 1.8 & -8.24  & 0.5 & 1.4         \vspace{1mm} \\
DE Tau &    M2     &   3560  &  1.01 & 0.07 &  1    & 2.6 & -7.69  & 0.4 & 0.2       \vspace{1mm} \\
DL Tau &    K7     &   4060  &  0.89 & 0.23 &  2    & 1.9 & -7.66  & 0.8 & 2.2           \vspace{1mm} \\
DM Tau &    M1     &   3705  &  0.24 & 0.10 &  2    & 1.2 & -7.99  & 0.5 & 3.6         \vspace{1mm}   \\
DN Tau &    M0     &   3850  &  0.91 & 0.02 &  1    & 2.2 & -8.55  & 0.6 & 1.1         \vspace{1mm}   \\
DO Tau &    M0     &   3850  &  0.79 & 0.60 &  1    & 2.0 & -7.10  & 0.6 & 1.3  \vspace{1mm} \\
DR Tau  &   K7     &   4060  &  1.09 & 0.72 &  2    & 2.1 & -7.12  & 0.8 & 1.5          \vspace{1mm} \\
DS Tau  &   K2     &   4900  &  0.90 & 0.21 &  1    & 1.3 & -8.00  & 1.1 & 17.0    \vspace{1mm}  \\
FM Tau  &   M0     &   3850  &  0.26 & 0.10 &  2    & 1.2 & -8.13  & 0.6 & 6.3       \vspace{1mm}  \\
FZ Tau &    M0     &   3850  &  1.06 & 0.46 &  3    & 2.3 & -7.14  & 0.6 & 0.8      \vspace{1mm} \\
GM Aur &    K3     &   4730  &  1.26 & 0.07 &  1    & 1.7 & -8.44  & 1.2 & 7.2     \vspace{1mm}  \\
GO Tau &    M0     &   3850  &  0.27 & 0.01 &  3    & 1.2 & -9.14  & 0.6 & 4.8         \vspace{1mm}  \\
HO Tau &   M0.5    &   3778  &  0.16 & 0.01 &  2    & 0.9 & -9.13  & 0.5 & 8.7        \vspace{1mm}  \\
IQ Tau &   M0.5    &   3778  &  0.73 & 0.04 &  2    & 2.0 & -8.24  & 0.6 & 1.2        \vspace{1mm}  \\
RY Tau &    K1     &   5080  &  6.59 & 1.60 &  4    & 3.3 & -7.03  & 2.2 & 1.1       \vspace{1mm}    \\
SU Aur &    G2     &   5860  &  10.6 & 0.55 &  4    & 3.2 & -7.52  & 1.7 & 2.2        \vspace{1mm}   \\
UZ Tau E &  M1     &   3705  &  2.27 & 0.31 &  2    & 3.7 &
-7.02  & 0.5 & 0.1          \vspace{2mm}
\\

\hline
\end{tabular}

\label{tab:sample}
\begin{flushleft}
$a$) References. (1) Gullbring et al.~(\cite{Gul98}); (2) Valenti et al.~(\cite{Val93}), these accretion luminosities have been multiplied by a factor $(140$ pc$/160$ pc$)^2$ to account for the current estimate of 140 pc for the distance to the Taurus-Auriga star forming region; (3) Herczeg et al.~(\cite{Her09}); (4) Calvet et al.~(\cite{Cal04}).
\end{flushleft}

\end{table*}

\subsection{Mass accretion rates}
\label{sec:mdot}

Given the stellar luminosities and effective temperatures we
calculated the stellar radii using the Stefan-Boltmann law. Once
the stellar mass $M_{\star}$ and radius $R_{\star}$ are set, the
mass accretion rate $\dot{M}_{\rm{acc}}$ can be directly obtained
from an estimate of the accretion luminosity $L_{\rm{acc}}$ by

\begin{equation}\label{eq:Macc}
 \dot{M}_{\rm{acc}}=1.25\frac{R_{\star} L_{\rm{acc}}}{G M_{\star}},
\end{equation}
where $G$ is the gravitational constant and the factor of $\sim1.25$
is calculated assuming a disk truncation radius of $\sim 5 \
R_{\star}$ (Gullbring et al.~\cite{Gul98}).

The estimates of $L_{\rm{acc}}$ were obtained from spectroscopic detections of
excess Balmer continuum emission in the literature (see references in Table \ref{tab:sample}).  These estimates are obtained primarily from measuring the excess flux in the Balmer continuum shortward of 3646 \AA.  The luminosity is then obtained by applying a bolometric correction calculated from shock models (Calvet \& Gullbring~\cite{Cal98}, Gullbring et al.~\cite{Gul00}) or simplistic plane-parallel slabs (Valenti et al.~\cite{Val93}, Gullbring et al.~\cite{Gul98}).

Two sources in our sample, FZ Tau and GO Tau, do not have literature estimates of accretion luminosity from the excess Balmer continuum emission.  We obtained the accretion luminosity for these two sources from low-resolution spectra from 3200--8700 \AA\ obtained with the Double Spectrograph at the 5m Hale Telescope at Palomar.   The accretion luminosity was then calculated from the excess Balmer continuum emission, following the method described in Herczeg \& Hillenbrand~(\cite{Her08}).

\subsection{Disks sub-mm and mm data from the literature}
\label{sec:mm-data}

In order to probe the sub-millimetre and millimeter SED we collected data of
the dust continuum emission between  $\sim 0.450$ to $\sim 7$ mm from several works in the literature, listed in Table
\ref{tab:literature_obs}.  These data, except for $7$ mm measurements, have been used together with the new PdBI observations at
$\sim 3$ mm to constrain the
disk properties by fitting the sub-mm/mm SED with the models
described in Section \ref{sec:models}. The $7$-mm data is excluded because free-free emission from ionized gas contaminates disk emission, with free-free emission typically contributing $\sim20$\% of the 7mm flux (Rodmann et al.~\cite{Rod06}).  Simultaneous cm observations are required to constrain the free-free emission at 7mm, which is variable on short timescales ($\simless$ a few days, see Lommen et al.~\cite{Lom09}). Only 2 disks in our sample (RY Tau and UZ Tau E) have been observed nearly simultaneously at centimeter and millimeter wavelengths (Rodmann et al.~\cite{Rod06}), without which the 7mm fluxes should be considered only as upper limits for the dust emission.  We therefore did not include the 7mm data in our analysis, although we included the available 7mm data from the literature in Figure \ref{fig:SED_fits} and verified a posteriori that the fluxes of our disk models for all sources are equal or less than the measured flux at 7mm.  Simultaneous 7mm and cm-wavelength observations with the EVLA will extend the coverage of the disk SED to longer wavelengths, in order to better constrain the millimeter-wavelengths dust opacity and to provide information on dust grains larger than $\sim$ 1 cm.

\begin{table*}
\centering \caption{ Literature sources for the sub-mm/mm data. } \vskip 0.1cm
\begin{tabular}{cc}
\hline \hline
\\
$\lambda$ (mm) & Source References$^{a}$ \vspace{1mm}
\\
\hline
\\

0.450  &  1          \vspace{1mm} \\
0.600  &  2, 3     \vspace{1mm} \\
0.624  &  4          \vspace{1mm} \\
0.769  &  4          \vspace{1mm} \\
0.800  &  2, 3     \vspace{1mm} \\
0.850  &  1, 5      \vspace{1mm} \\
0.880  &  5         \vspace{1mm}  \\
1.056  &  4         \vspace{1mm}  \\
1.100  &  2, 3     \vspace{1mm}  \\
1.200  &  6     \vspace{1mm}  \\
1.300  &  1, 7, 8, 9, 10     \vspace{1mm}  \\
1.330  &  5     \vspace{1mm}  \\
2.000  &  2, 11     \vspace{1mm}  \\
2.700  &  7, 8, 12, 13     \vspace{1mm}  \\
3.100  &  8     \vspace{1mm}  \\
3.400  &  9, 14     \vspace{1mm}  \\
7.000  &  15
\vspace{2mm}
\\

\hline
\\
\end{tabular}

\label{tab:literature_obs}
\begin{flushleft}
$a$) References. (1) Andrews \& Williams~(\cite{And05}); (2) Mannings \& Emerson~(\cite{Man94}); (3) Adams et al.~(\cite{Ada90}); (4) Beckwith \& Sargent~(\cite{Bec91}); (5) Andrews \& Williams~(\cite{And07}); (6) Althenoff et al.~(\cite{Alt94}); (7) Dutrey et al.~(\cite{Dut96}); (8) Jensen et al.~(\cite{Jen96}); (9) Koerner et al.~(\cite{Koe95}); (10) Isella et al.~(\cite{Ise09}); (11) Kitamura et al.~(\cite{Kit02}); (12) Mundy et al.~(\cite{Mun96}); (13) Schaefer et al.~(\cite{Sch09}); (14) Ohashi et al.~(\cite{Oha96}); (15) Rodmann et al.~(\cite{Rod06}).
\end{flushleft}
\end{table*}

\section{Disk models}
\label{sec:models}
We compare the observational sub-mm/mm data with the two-layer
models of flared disks (i.e. in hydrostatic equilibrium) heated by
the stellar radiation as developed by Dullemond et al.~(\cite{Dul01}), following the
schematization of Chiang \& Goldreich~(\cite{Chi97}). These models have been used in the
analysis of CQ Tau by Testi et al.~(\cite{Tes03}) and of a sample of 9 Herbig Ae
stars by Natta et al.~(\cite{Nat04}) and we refer to these papers for a more
detailed description.

To completely characterize a model of the disk, we need to specify
the stellar properties (i.e. the luminosity $L_{\star}$, effective
temperature \Tstar, mass $M_{\star}$ and distance $d$, assumed to
be $140$ pc for all the sources in our sample), some characteristics of the disk
structure, namely the inner and outer radius ($R_{\rm{in}}$ and
$R_{\rm{out}}$, respectively), the parameters $\Sigma_{1}$ and $p$
defining the radial profile of the dust surface density assumed to
be a power-law\footnote{As shown by Hughes et al.~(\cite{Hug08}) and Isella et al.~(\cite{Ise09}), a tapered exponential edge in the surface density profile, physically motivated by the viscous evolution of the disk, is able to explain the apparent discrepancy between gas and dust outer radii derived from millimeter observations of protoplanetary disks. Here we want to note that when fitting the sub/mm-mm SED with an exponential tail instead of a truncated power-law, the dust opacity spectral index $\beta$ (defined in Section \ref{sec:opacity}) is unchanged, and disk masses vary by less than a factor of 2, even if the outer disk radius becomes much bigger (see, e.g., Table 4 in Kitamura et al.~\cite{Kit02}). So the simpler truncated power-law model is a justified approximation for the purposes of this paper.} $\Sigma_{\rm{dust}} = \Sigma_{\rm{dust},1} (R/1$
AU$)^{-p}$, the disk inclination angle $i$  ($90$\degree \ for an
edge-on disk) and, lastly, the dust opacity.
The millimeter SED is almost completely insensitive to some of the disk
parameters, which cannot thus be constrained by our analysis.
For example, the inner radius of the disk, $R_{\rm{in}}$, affects the
emission only at near and mid-infrared wavelengths.

The outer radius $R_{\rm{out}}$ is a parameter that can be probed
by high-angular resolution interferometric observations. In our
sample 13 disks have been spatially resolved, yielding
an estimate for $R_{\rm{out}}$. It is important to note that
different interferometric observations performed on the same disk
can lead to very different values of $R_{\rm{out}}$ ($\ge 100$
AU) because of differences in the observations angular resolution, sensitivity,
and methods for deriving the radius.  Given these uncertainties in the determination of $R_{\rm{out}}$,
we consider a range of possible values for $R_{\rm{out}}$ rather than a single-value estimate, as listed in Table
\ref{tab:outer_radii}.
For the 8 objects in our sample that have not yet been mapped with angular resolutions smaller than a few arcseconds which may potentially resolve the disk, a fiducial interval of $100-300$ AU has been assumed\footnote{We will refer to these sources (CW, CX, DE, DS, FM, FZ, HO Tau, and SU Aur) as ``unmapped'' for the rest of the paper. Note that they are all faint with $F_{\rm{1mm}} < 120$ mJy. This probably explains why for these objects no observations at high-angular resolution have been attempted yet: the sensitivity of the current (sub-)mm interferometers would be not high enough to detect the low surface brightness of the outer disk regions.}. This adopted range for $R_{\rm{out}}$ comprises the vast majority of the disk outer radii
distribution function as derived by Andrews \& Williams~(\cite{And07}) through $\sim 1''$ angular resolution sub-millimeter observations of 24 disks in Taurus-Auriga and $\rho-$Ophiucus. We will discuss the impact of the outer
disk uncertainty on the results of our analysis in Section
\ref{sec:results}. Here we want to note that with these values of
$R_{\rm{out}}$ the disk emission at sub-mm/mm wavelengths is dominated by the outer disk regions ($R
> 30$ AU) which are optically thin to their own radiation. An
effect of this is that the millimeter SED does not depend on the disk
inclination angle $i$ (for $i \simless 80^{\circ}$), thus decreasing the number of the model
parameters used to fit the observational SED. This is
certainly true for the spatially resolved disks for which we have an
estimate for $R_{\rm{out}}$. The 8
unmapped objects could have optically thick
inner disks that contribute significantly to the
sub-mm/mm emission.  Since the disk outer radius cannot be
constrained by the SED alone, interferometric imaging that
spatially resolves these disks is the only way to get information
on the exact value of $R_{\rm{out}}$. Nevertheless, since the
disks that have been spatially resolved so far show outer radii
typically in the range $100 - 300$ AU, we consider highly unlikely that a
significant fraction of the unmapped disks have
$R_{\rm{out}} \simless 50$ AU, which would make the
disk optically thick even at mm wavelengths.

\subsection{Dust opacity}
\label{sec:opacity}

The last quantity that defines the disk model is the dust opacity, which
depends on grain sizes, chemical compositions, and shapes (e.g. Miyake \& Nakagawa~\cite{Miy93},
Pollack et al.~\cite{Pol94}, Draine et al.~\cite{Dra06}). In order to extract from the sub-mm/mm SED quantitative estimates for the dust grain
sizes and the dust mass in the disks, assumptions on the chemical
composition and shape of the dust grains have to be made.

\begin{figure*}[htb!]
 \centering
 \begin{tabular}{cc}
 \includegraphics[scale=0.54]{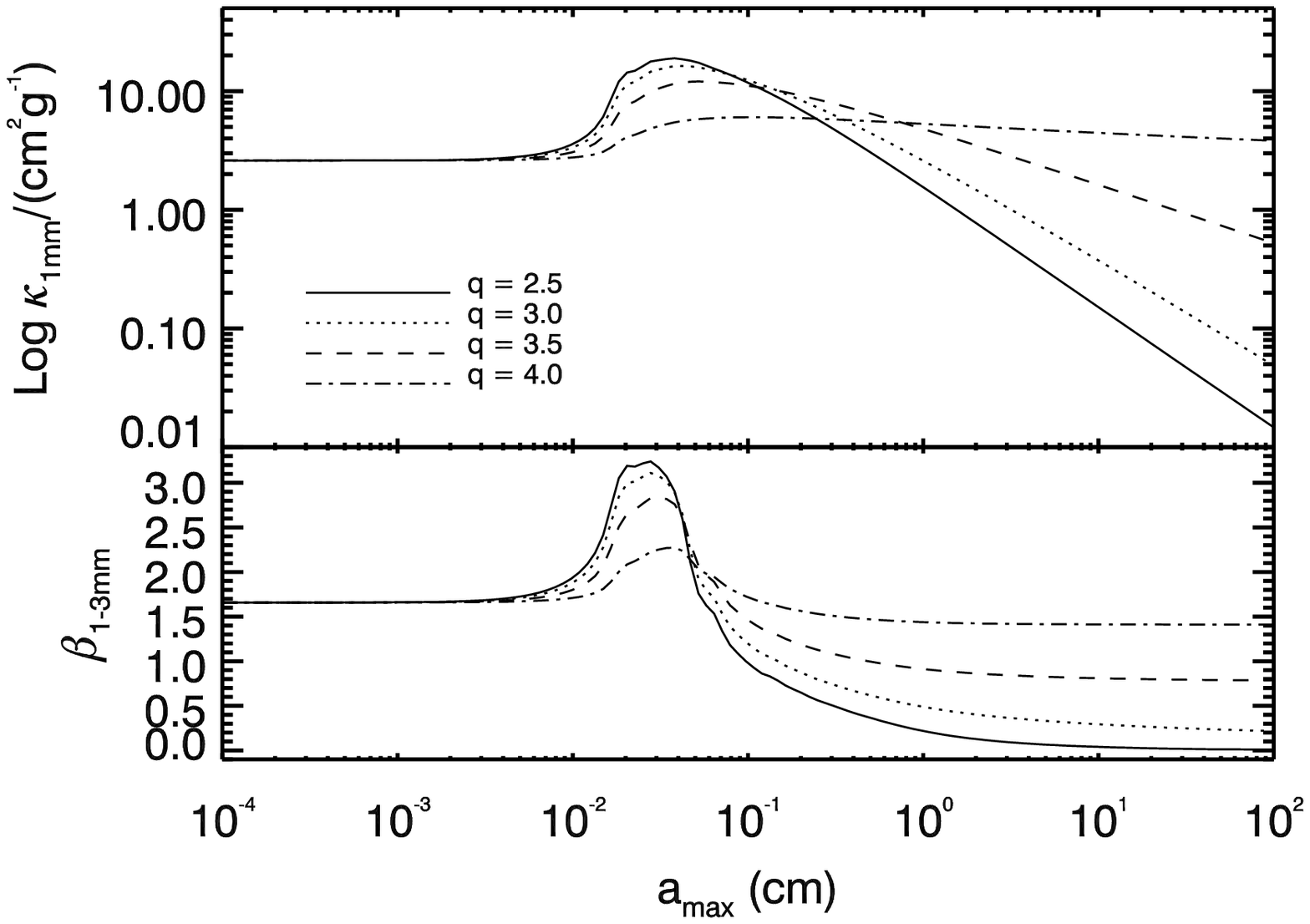} &
 \includegraphics[scale=0.45]{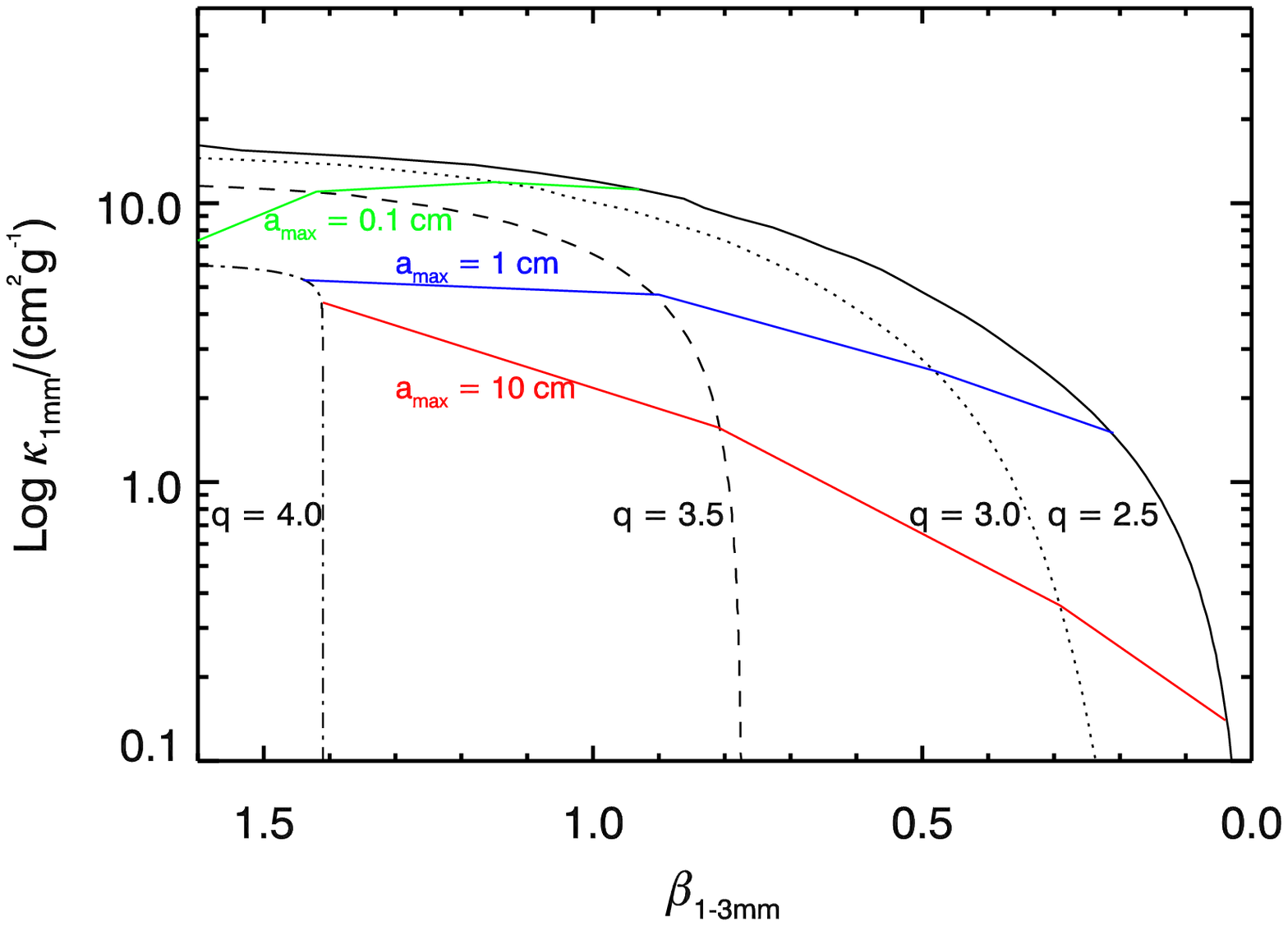} \\

 \end{tabular}
 \caption{Millimeter dust opacity for the adopted dust model. Left top panel: dust opacity at 1 mm as function of $a_{\rm{max}}$ for a grain
 size distribution $n(a) \propto a^{-q}$ between $a_{\rm{min}}=0.1\ \mu$m and $a_{\rm{max}}$.
 Different curves are for different values of $q$, as labelled. The dust grains adopted here are porous composite spheres made of astronomical silicates ($\approx 10\%$ in volume), carbonaceous materials ($\approx 20\%$) and water ice ($\approx 30\%$; see text for the references to the optical constants). Left bottom panel: $\beta$ between $1$ and $3$ mm as a
 function of $a_{\rm{max}}$ for the same grain distributions. Right panel: dust opacity at 1 mm as function
 of $\beta$ between $1$ and $3$ mm for the same grain distributions; different iso-$a_{\rm{max}}$ curves for $a_{\rm{max}} = 0.1, 1, 10$ cm are shown. In the plots the unit of measure for the dust opacity is cm$^2$ per gram of dust.}

 \label{fig:dust}

\end{figure*}

For our discussion we considered porous composite spherical grains
made of astronomical silicates, carbonaceous materials
and water ice (optical constants for the individual components from Weingartner \& Draine~\cite{Wei01}, Zubko et al.~\cite{Zub96}, Warren~\cite{War84} respectively) adopting fractional abundances used by Pollack et al.~(\cite{Pol94}) (see caption of Figure \ref{fig:dust}). We considered a population of grains with a
power-law size distribution $n(a) \propto a^{-q}$ between a
minimum and a maximum size, $a_{\rm{min}}$ and $a_{\rm{max}}$,
respectively. In the ISM, $a_{\rm{min}}$ is a few tens of \AA,
$a_{\rm{max}} \sim 0.1-0.2\ \mu$m, and $q = 3.5$ (Mathis et al.~\cite{Mat77}, Draine \& Lee~\cite{Dra84}, by analysing extinction and scattering of starlinght from the interstellar dust). As discussed e.g. in
Natta et al.~(\cite{Nat04}) in a protoplanetary disk, because of grain
processing, one expects much larger values of
$a_{\rm{max}}$ than in the ISM, and a variety of possible $q$
values, depending on the processes of
grain coagulation and fragmentation that occur in the disk.
Experimental studies of fragmentation from a single target have found values of $q$ ranging from $\approx 1.9$ for low-velocity collisions to $\approx 4$ for catastrophic impacts (Davies \& Ryan~\cite{Dav90}), but the power-law index for fragments from shattering events integrated over a range of target masses need not be necessarily the same as in the case of a single target. In the disk, if the coagulation processes from the initial ISM-grains population dominate over the fragmentation ones, then a value for $q$ that is lower than the ISM one is expected (the opposite should occur in the case of fragmentation playing the major role).

In the adopted two-layer disk model, the disk is divided into the surface and midplane regions, and therefore two different opacity laws are considered.
Because of the assumed vertical hydrodynamical equilibrium and dust settling, in a protoplanetary disk the content of mass in the midplane is much higher than in the surface.  At millimeter wavelengths, where the disk is mostly optically thin to its own radiation, the midplane total emission dominates over the surface one (although the surface dust plays a crucial role for the heating of the disk). As a consequence, from our analysis we can extract information only on the dust opacity of the midplane, and for the rest of the paper when we consider the dust opacity we will mean only the midplane component.

For the dust grains in the disk surface we assumed the same chemical composition and shape as for the midplane grains, whereas for their size distribution we assumed for $a_{\rm{min}}$ and $a_{\rm{max}}$ values around $0.1 \ \mu$m (the sub-mm SED is insensitive to these values as long as $a_{\rm{max}}$ in the surface is much lower than $a_{\rm{max}}$ in the midplane, as expected if dust settling is important).

\begin{table*}
\centering \caption{ Adopted disks outer radii. } \vskip 0.1cm
\begin{tabular}{lcccc}
\hline \hline
\\
Object & $R_{\rm{out}}^{\rm{a}}$ (K02) & $R_{\rm{out}}^{\rm{a}}$ (AW07) & $R_{\rm{g}}^{\rm{b}}$ (I09) & Adopted $R_{\rm{out}}-$interval \\
            & (AU) & (AU) & (AU) & (AU)\\

\\
\hline
\\

AA Tau  & $214^{+63}_{-51}$ & $400^{+600}_{-75}$ & ...    & $200-400$         \vspace{1mm} \\
CI Tau  & ...      & $225\pm 50$        & ...    & $150-250$         \vspace{1mm} \\
CW Tau  & ...      & ...       & ...    & $100-300$         \vspace{1mm} \\
CX Tau  & ...      & ...       & ...    & $100-300$         \vspace{1mm} \\
CY Tau  & $211\pm 28$       & ...       & $230$           & $230-330$               \vspace{1mm} \\
DE Tau  & ...      & ...       & ...    & $100-300$         \vspace{1mm} \\
DL Tau  & $152^{+28}_{-27}$ & $175^{+50}_{-25}$  & ...    & $150-250$         \vspace{1mm} \\
DM Tau  & $220\pm 47$       & $150^{+250}_{-100}$ & $160$          & $160-260$         \vspace{1mm} \\
DN Tau  & $147\pm 469$      & $100^{+300}_{-25}$ & $125$           & $125-225$         \vspace{1mm} \\
DO Tau  & $98\pm 52$        & ...       & ...    & $100-200$         \vspace{1mm} \\
DR Tau  & $193\pm 141$      & $100^{+175}_{-25}$ & $90$            & $90-190$         \vspace{1mm} \\
DS Tau  & ...      & ...       & ...    & $100-300$         \vspace{1mm}  \\
FM Tau  & ...      & ...       & ...    & $100-300$         \vspace{1mm}  \\
FZ Tau  & ...      & ...       & ...    & $100-300$         \vspace{1mm} \\
GM Aur  & $151^{+63}_{-29}$ & $150\pm 25$        & $270$           & $270-370$      \vspace{1mm}  \\
GO Tau  & ...      & $350^{+650}_{-175}$ & $160$          & $160-260$         \vspace{1mm}  \\
HO Tau  & ...      & ...       & ...    & $100-300$         \vspace{1mm}  \\
IQ Tau  & $329^{+403}_{-70}$ & ...      & ...    & $250-400$         \vspace{1mm}  \\
RY Tau  & $81\pm 128$       & $150\pm 25$        & $115$           & $115-215$         \vspace{1mm}    \\
SU Aur  & ...      & ...       & ...    & $100-300$         \vspace{1mm}   \\
UZ Tau E  & ...    & ...       & $160$           & $160-260$      \vspace{2mm} \\

\hline
\\
\end{tabular}

\begin{flushleft}
a) $R_{\rm{out}}$ listed here results from image fitting using a truncated power-law for the surface density profile. K02: Kitamura et al.~(\cite{Kit02}); AW07: Andrews \& Williams~(\cite{And07}).
b) $R_{\rm{g}}$ is the radius within which the 95\% of the source flux is observed, after fitting the visibility of the source with a gaussian; we used this value, when available, as a lower limit for our adopted $R_{\rm{out}}$-interval. I09: Isella et al.~(\cite{Ise09}).
\end{flushleft}

\label{tab:outer_radii}
\end{table*}

Once the chemical composition and
shape of the dust grains in the midplane are set, the dust opacity law depends on
$a_{\rm{min}}$, $a_{\rm{max}}$ and $q$.
Considering a value of $0.1 \ \mu$m for $a_{\rm{min}}$ (the dependence of the millimeter dust opacity from this parameter turns out to be very weak), the models parameters for our analysis are the stellar parameters ($L_{\star}$, $T_{\rm{eff}}$, $M_{\star}$), that have been set and listed in Table \ref{tab:sample}, plus the disk ones
($\Sigma_{\rm{dust},1}$, $p$, $R_{\rm{out}}$, $q$, $a_{\rm{max}}$), or
equivalently ($M_{\rm{dust}}$, $p$, $q$, $a_{\rm{max}}$, $R_{\rm{out}}$)
where $M_{\rm{dust}}$ is the mass of dust in the disk.

For reasons described in Section \ref{sec:results} it is convenient to
approximate the millimeter dust opacity law discussed so far in terms of a power-law at millimiter wavelengths $\kappa=\kappa_{\rm{1 mm}}$($\lambda/$1 mm)$^{-\beta}$, with $\kappa_{\rm{1 mm}}$ in
units of cm$^2$ per gram of dust. At a fixed value of $q$ the relations between $\kappa_{\rm{1 mm}}$, $\beta$ and $a_{\rm{max}}$ can be determined. These are shown in Figure \ref{fig:dust} for $q=2.5,3,3.5,4$.

\section{Results}
\label{sec:results}
\begin{figure*}
 \centering
 \includegraphics[scale=0.7]{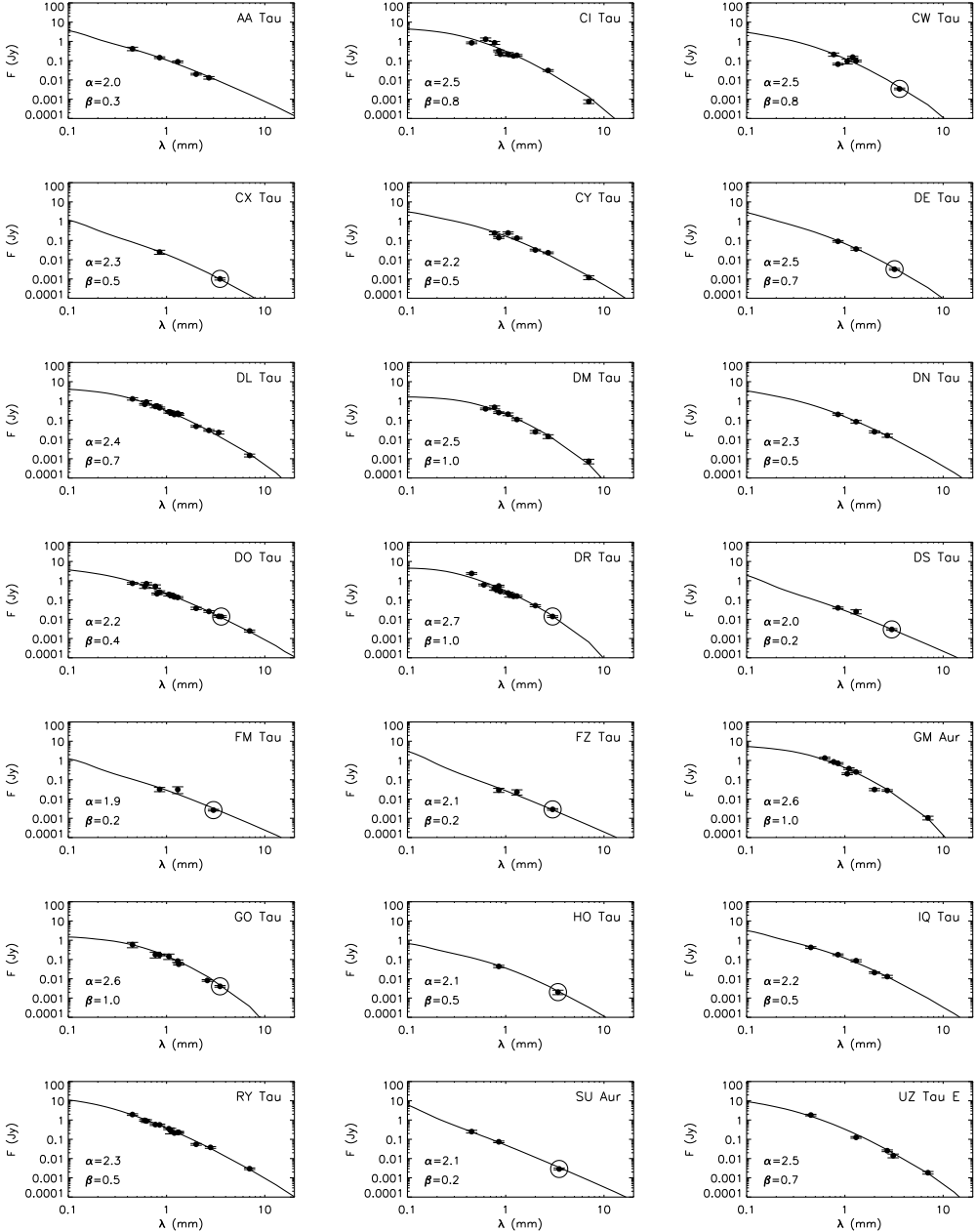}
 \caption{Fits of the sub-mm/mm SEDs with the two-layer flared disk models (solid lines). Note that even if the data at 7 mm, available only for 8 disks in our sample, are present in the plots, they have not been used in the fitting procedure (see Section \ref{sec:mm-data}). The errorbars in the plots take into account an uncertainty of 10\% on the absolute flux scale, typical for flux measurements at these wavelengths. The circled data come from the new PdBI observations. The fitting values of the spectral index $\alpha$ and of the dust opacity spectral index $\beta$ between 1 and 3 mm are indicated in the bottom left corner of each plot. The value of the surface density power-law index $p$ adopted in these fits is 1, and the adopted value for the outer disk radius is listed in Table \ref{tab:fits_results}.}
 \label{fig:SED_fits}
\end{figure*}

The sub-mm/mm SED fits for all the objects in our sample are
reported in Figure \ref{fig:SED_fits}\footnote{We did not include existing NIR and mid-IR data in our analysis because at these shorter wavelengths the disk emission is optically thick and thus depends on the properties of the dust grains located in the surface layers of the inner disk regions. The aim of our analysis is instead to probe the optically thin disk emission in order to investigate the dust properties in the disk midplane.}. As discussed in
Testi et al.~(\cite{Tes03}) and Natta et al.~(\cite{Nat04}), the only quantities that can be
constrained by the sub-mm/mm SED are the millimeter dust opacity spectral
index $\beta$ (in this paper calculated between 1 and 3 mm) and the $M_{\rm{dust}} \times
\kappa_{\rm{1 mm}}$ product\footnote{Note that, according to the adopted disk model, after fixing the stellar parameters, and these two quantities from the observed SED, the midplane temperature is well constrained.}. From the SED alone it is impossible to constrain the value of the surface density exponent $p$, since
it is generally possible to fit the SED data with either very flat
($p = 0.5$) or very steep ($p = 2$) surface density profiles.
However, all available high angular resolution observations
performed so far suggest that $p \leq 1.5$ (e.g. Dutrey et al.~\cite{Dut96},
Wilner et al.~\cite{Wil00}, Kitamura et al.~\cite{Kit02}, Testi et al.~\cite{Tes03}, Andrews \& Williams~\cite{And07}, Isella et al.~\cite{Ise09}, Andrews et al.~\cite{And09}).

The degeneracy of the SED on $p$ and $R_{\rm{out}}$ has an impact onto the derived uncertainties for
$\beta$ and $M_{\rm{dust}} \times
\kappa_{\rm{1 mm}}$. Considering both this degeneracy and uncertainties of the observational data, the total absolute uncertainties are
approximately $0.2-0.3$ for $\beta$ and a factor of $\approx 3$
and $\approx 4$ for $M_{\rm{dust}} \times
\kappa_{\rm{1 mm}}$ for the spatially resolved and unmapped disks respectively. These uncertainties are mainly due to the adopted ranges for $R_{\rm{out}}$ (listed in Table \ref{tab:outer_radii}) and $p$ between 0.5 and 1.5, whose values determine the impact of the optically thick regions of the disk to the total emission, as explained in Section \ref{sec:models}.

The estimates for $\beta$ and $M_{\rm{dust}} \times
\kappa_{\rm{1 mm}}$ obtained by
our analysis are reported in Table \ref{tab:fits_results}.

\begin{table*}[htb!]
\centering \caption{ Disk properties. } \vskip 0.1cm
\begin{tabular}{lrccccccc}
\hline \hline
\\
Object name$^{\rm{a}}$ & $R_{\rm{out}}^{\rm{b}}$ & $F_{\rm{1mm}}^{\rm{c}}$ & $\alpha_{\rm{1-3}mm}$  & $\beta_{\rm{1-3}mm}$ & $M_{\rm{dust}} \times
\kappa_{\rm{1 mm}}$ & $M_{\rm{dust}}^{q=2.5}$ & $M_{\rm{dust}}^{q=3}$ & $M_{\rm{dust}}^{q=3.5}$    \vspace{1mm} \\
           & (AU) & (mJy)  &              &         &  ($M_{\odot} \times$ cm$^2$g$^{-1}$) & ($M_{\odot}$) & ($M_{\odot}$) & ($M_{\odot}$)\\

\\
\hline
\\

AA Tau  & 300 & 108 & 2.0 &  0.3  & $2.8\cdot10^{-4}$ & $1.5\cdot10^{-4}$ & $1.5\cdot10^{-3}$ & ...          \vspace{1mm} \\
CI Tau  &  200 & 314 & 2.5   &  0.8  & $7.9\cdot10^{-4}$ & $1.1\cdot10^{-4}$ & $1.2\cdot10^{-4}$ & $5.0\cdot10^{-4}$          \vspace{1mm} \\
\underline{CW Tau}              &  200 & 129 & 2.5  &  0.8  & $3.2\cdot10^{-4}$ & $4.2\cdot10^{-5}$ & $4.4\cdot10^{-5}$ & $1.9\cdot10^{-4}$          \vspace{1mm} \\
\underline{CX Tau}              &  200 & 19 & 2.3  &  0.5  & $5.7\cdot10^{-5}$ & $1.2\cdot10^{-5}$ & $1.9\cdot10^{-5}$ & ...      \vspace{1mm} \\
CY Tau  &  280 & 168 & 2.2  &  0.5  & $6.0\cdot10^{-4}$ & $1.3\cdot10^{-4}$ & $2.0\cdot10^{-4}$ & ...          \vspace{1mm} \\
\underline{DE Tau}              &  200 & 69 & 2.5  &  0.7  & $1.6\cdot10^{-4}$ & $2.4\cdot10^{-5}$ & $2.8\cdot10^{-5}$ & ...      \vspace{1mm} \\
DL Tau  &  200 & 313 & 2.4  &  0.7  & $8.4\cdot10^{-4}$ & $1.3\cdot10^{-4}$ & $1.5\cdot10^{-4}$ & ... \vspace{1mm} \\
DM Tau  &  210 & 209 & 2.5  &  1.0  & $9.2\cdot10^{-4}$ & $1.0\cdot10^{-4}$ & $1.0\cdot10^{-4}$ & $1.6\cdot10^{-4}$    \vspace{1mm} \\
DN Tau  &  175 & 153 & 2.3  &  0.5  & $3.5\cdot10^{-4}$ & $7.7\cdot10^{-5}$ & $1.2\cdot10^{-4}$ & ... \vspace{1mm} \\
DO Tau  &  150 & 220 & 2.2  &  0.4  & $4.9\cdot10^{-4}$ & $1.4\cdot10^{-4}$ & $3.1\cdot10^{-4}$ & ... \vspace{1mm} \\
DR Tau  &  140 & 298 & 2.7  &  1.0  & $6.6\cdot10^{-4}$ & $7.4\cdot10^{-5}$ & $7.2\cdot10^{-5}$ & $1.2\cdot10^{-4}$          \vspace{1mm} \\
\underline{DS Tau}              &  200 & 28 & 2.0  &  0.2  & $6.2\cdot10^{-5}$ & $4.0\cdot10^{-5}$ & $1.1\cdot10^{-3}$ & ...  \vspace{1mm}  \\
\underline{FM Tau}              &  200 & 29 & 1.9  &  0.2  & $6.5\cdot10^{-5}$ & $6.0\cdot10^{-5}$ & $1.7\cdot10^{-3}$ & ... \vspace{1mm}  \\
\underline{FZ Tau}              & 200 & 27 &  2.1  &  0.2  & $5.0\cdot10^{-5}$ & $3.2\cdot10^{-5}$ & $9.0\cdot10^{-4}$ & ... \vspace{1mm} \\
GM Aur  &  320 & 423 & 2.6  &  1.0  & $1.4\cdot10^{-3}$ & $1.6\cdot10^{-4}$ & $1.5\cdot10^{-4}$ & $2.5\cdot10^{-4}$          \vspace{1mm}  \\
GO Tau  &  210 & 151 & 2.6  &  1.0  & $6.3\cdot10^{-4}$ & $7.1\cdot10^{-5}$ & $6.9\cdot10^{-5}$ & $1.1\cdot10^{-4}$          \vspace{1mm}  \\
\underline{HO Tau}              &  200 & 36 & 2.1  &  0.5  & $1.7\cdot10^{-4}$ & $3.7\cdot10^{-5}$ & $5.6\cdot10^{-5}$ & ... \vspace{1mm}  \\
IQ Tau  &  325 & 118 & 2.2  &  0.5  & $3.4\cdot10^{-4}$ & $7.4\cdot10^{-5}$ & $1.1\cdot10^{-4}$ & ... \vspace{1mm}  \\
RY Tau  &  165 & 383 & 2.3  &  0.5  & $5.1\cdot10^{-4}$ & $1.1\cdot10^{-4}$ & $1.7\cdot10^{-4}$ & ... \vspace{1mm}    \\
\underline{SU Aur}              &  200 & 50 & 2.1  &  0.2  & $5.0\cdot10^{-5}$ & $3.2\cdot10^{-5}$ & $9.0\cdot10^{-4}$ & ... \vspace{1mm}   \\
UZ Tau E & 210 & 333 & 2.5  &  0.7  & $6.1\cdot10^{-4}$ & $9.6\cdot10^{-5}$ & $1.1\cdot10^{-4}$ & ... \vspace{2mm} \\

\hline
\\
\end{tabular}

\begin{flushleft}
a) Underlined objects are those that have not been mapped to date through high-angular resolution imaging.
b) Central value of the adopted $R_{\rm{out}}$-interval reported in Table \ref{tab:outer_radii}. This is the value used to extract the other quantities listed in the Table.
c) Source flux density at 1 mm from the best fit disk model described in Section \ref{sec:models}.
\end{flushleft}

\label{tab:fits_results}
\end{table*}

\subsection{Spectral slopes and dust opacity index}
\label{subsec:spectral_slopes}

\begin{figure}
 \centering
 \resizebox{\hsize}{!}{\includegraphics{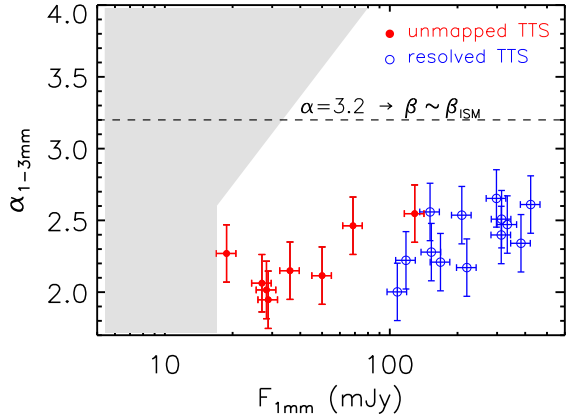}}
 \caption{Spectral index $\alpha$ between 1 and 3 mm plotted against the flux at 1 mm from the model best fits shown in Figure \ref{fig:SED_fits}. The points plotted here are listed in the first two columns of Table \ref{tab:fits_results}. Open blue points represent the spatially resolved disks, whereas the filled red points are for the unmapped ones. A value of $\alpha$ consistent with a disk model with ISM-like dust and an outer radius of $\sim 100$ AU is indicated as a dashed horizontal line. The grey region shows the area in the plot that is unexplored because of the $3\sigma$ sensitivity limits of our new and past millimeter observations (here considered as 15 mJy and 1.0 mJy at 1 and 3 mm respectively).}
 \label{fig:flux_alpha}
\end{figure}

Before analysing the values of $\beta$, which provide information on the level of grain growth in the outer disk regions, it is important to check whether our results could be affected by detection biases: indeed, for a given flux at $\sim 1$mm, a disk with large dust grains in the outer parts will produce a stronger emission at $\sim 3$ mm than a disk with a population of smaller-sized grains and be easier to detect. In Figure \ref{fig:flux_alpha} the grey region of the plot indicates the area in the ($F_{\rm{1mm}}$, $\alpha_{\rm{1-3mm}}$) plane in which the observations carried out so far are not sensitive, due to the sensitivity limits obtained at 1 and 3 mm. While
we could investigate the dust properties for 4 disks with $15 \ \rm{mJy} < F_{\rm{1mm}} < 30 \ \rm{mJy}$, the plot shows that a disk with a 1 mm flux in this same range, but with a spectral index $\alpha_{\rm{1-3mm}} \simgreat 3.2$, consistent with a disk model with ISM-like dust and an outer radius of $\sim 100$ AU, would not be detected at 3 mm. Nevertheless, we want to point out that only 2 of our 16 targets with $F_{\rm{1mm}} < 100$ mJy were not detected at 3mm, and both have very weak $F_{\rm{1.3mm}}$ fluxes (upper limits of $27$ mJy and $21$ mJy for DP Tau and GK Tau respectively, Andrews \& Williams~\cite{And05}), therefore the lack of disks with $\alpha \sim 3.2$ at $F_{\rm{1mm}} > 30$ mJy does not seem to be due to our sensitivity limit but is most likely real. Furthermore, as evident from the same plot, the 6 faintest disks, with $F_{\rm{1mm}} < 60 \ \rm{mJy}$, show a spectral index between 1 and 3 mm that is lower on average ($2.08 \pm 0.08$) than for the brighter disks ($2.40 \pm 0.06$). This is the reason of the discrepancy of the average value of the dust opacity spectral index $<\beta>$ between unmapped and resolved disks (see discussion below).

In Figure \ref{fig:dust} we plot $\kappa_{\rm{1mm}}$ and $\beta$
as functions of $a_{\rm{max}}$ for different values of $q$. One interesting result is
that the values of $\beta$
obtained for the objects in our sample by the SED-fitting
procedure are all less than or equal to $1$, which can be explained only if
dust grains with sizes larger than $1$ mm are present in the disk.
For the 9 sources with $\beta \leq 0.5$, the largest dust grains
must have grown to sizes $\ge 1$ cm according to our dust model.
If we let $q$ assume values lower than $2.5$, thus increasing the
fractional number of the larger grains, the maximum grain size
decreases, but $a_{\rm{max}} \simgreat 1$ mm for $\beta \simless
1$ is still needed.

The estimates of the maximum grain sizes $a_{\rm{max}}$ depend on
the dust model that one adopts. If we consider for example the
dust model adopted in Natta et al.~(\cite{Nat04}) (5\% olivine, 15\% organic
materials, 35\% water ice and 50\% vacuum), the estimates for
$a_{\rm{max}}$ are greater by about an order of magnitude than the
ones obtained with our dust model. In general, no realistic dust
grain models result in $\beta < 1$ for $a_{\rm{max}} < 1$ mm.

\begin{figure}
 \centering
 \resizebox{\hsize}{!}{\includegraphics{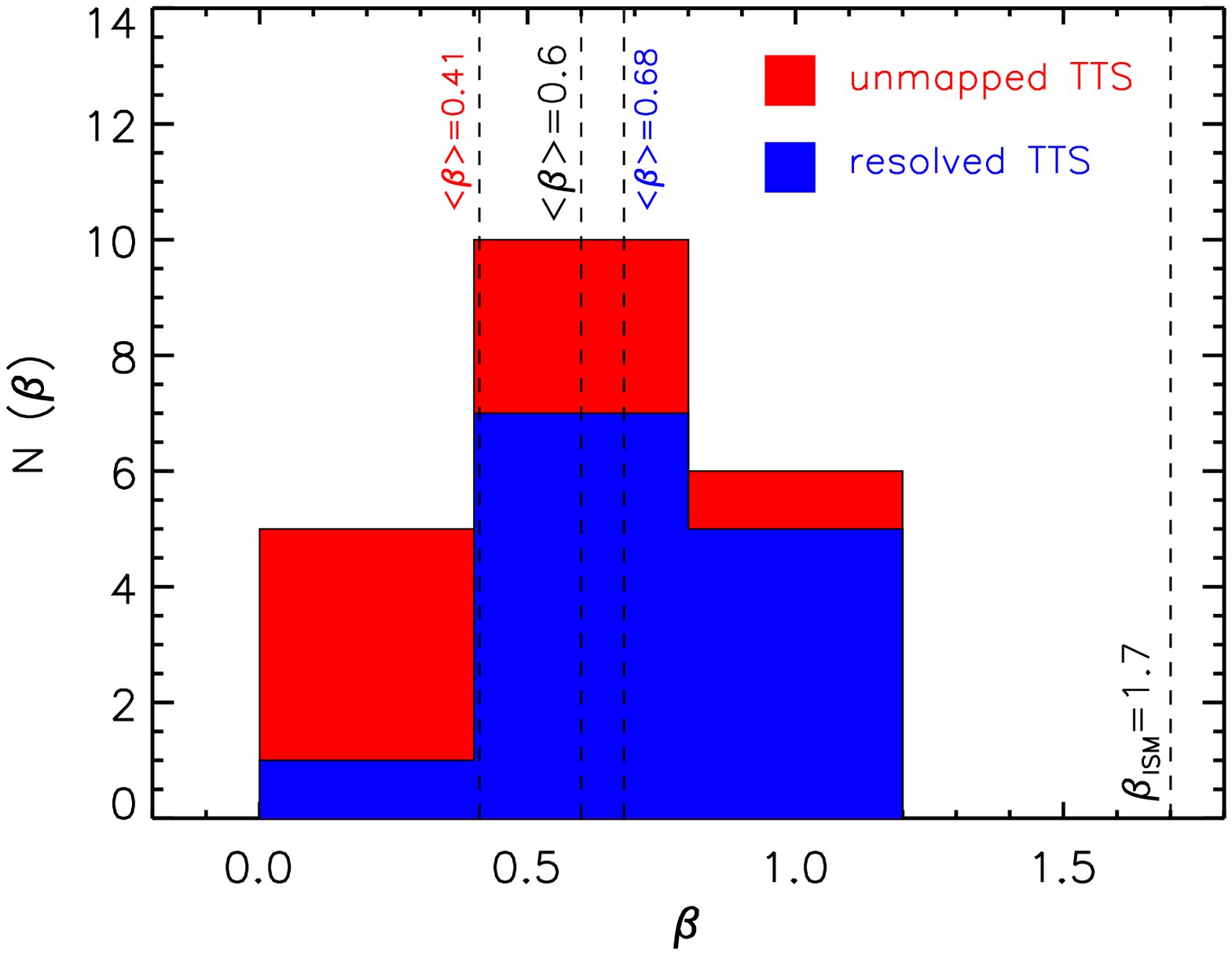}}
 \caption{Distribution of the dust opacity spectral index $\beta$ for the disks in our sample. In blue the values for the spatially resolved disks are indicated, whereas the red is for the unmapped ones. The average $\beta$ value for all the sources in our sample (black), for the unmapped sources only (red), for the resolved sources only (blue), and the value of 1.7 for the ISM dust are indicated as dashed vertical lines.}
 \label{fig:beta_hist}
\end{figure}

Here
it is worthwhile to remember that, whereas for the 13 resolved disks in our sample $\beta$ is well determined, for the 8 disks for which we do not have information on their spatial extension,
the optically thick parts of a possible compact disk could
significantly contribute to the emission.  If that is the case,
no conclusions on the dust grain properties could be drawn. In
particular, for the 6 unmapped disks with $\alpha
\approx 1.9-2.2$ (CX Tau, DS Tau, FM Tau, FZ Tau, HO Tau, SU Aur)
a very compact disk with $R_{\rm{out}} \approx 20$ AU, a relative
high inclination $i \approx 70^{o}$, and with ISM-like dust grains
would fit the sub-mm/mm SED. For the other 2 unmapped sources, DE
Tau ($\alpha \approx 2.4$) and CW Tau ($\alpha \approx 2.5$), ISM-like dust grains are
consistent with the sub-mm/mm data using a disk model with $i
\approx 70^{o}$ and $R_{\rm{out}} \approx$ 30, 40 AU
respectively. For the faintest disks, the outer
radius has to be small because if the disks were larger they
would also need to be more massive in order to keep being
optically thick, thus producing an $\alpha$ value close to 2 with
all the possible dust properties.  In that case, they would emit more
sub-mm/mm-wave radiation than observed. Since, for an optically
thick disk, decreasing the inclination angle has the effect of
making the SED steeper (because the central, hotter parts are less
obscured), for disks with $i < 70^{o}$ the outer radii would need
to be even smaller to fit the SED. Since these disk outer radii
are lower by factors of $2-4$ than the smallest disks that have
been spatially resolved so far, and since for all the 13 resolved
disks the values of $\beta$ have been found always much lower than
in the ISM ($\beta_{\rm{ISM}} \approx 1.7$, see Figure
\ref{fig:beta_hist}), thus showing clear evidence of grain growth,
we consider very unlikely that many of the unmmapped
sources are compact, optically-thick disks with unprocessed ISM
grains. For AA Tau, which is the faintest of the mapped sources ($F_{\rm{1mm}} \approx 100$ mJy) and has a low spectral index of $\alpha \approx 2.0$ typical of fainter sources, the maps indicate a disk radius of $200-400$ AU ruling out the possibility of optically thick disk regions dominating the total emission, and requiring a low $\beta$ of 0.3 to explain the shallow slope.

Also, only for the disks with $\beta \simgreat 0.8$, values of $q
\simgreat 3.5$ are consistent with the data (see right panel in Figure \ref{fig:dust}), in which case
$a_{\rm{max}} \geq 10$ cm. This is in good agreement with the
following consideration using the approximated asynthotic formula
$\beta \approx (q - 3)\beta_{RJ}$, derived by Draine~(\cite{Dra06}), where
$\beta_{\rm{RJ}}$ is the opacity spectral index of the solid
material in the Rayleigh limit, and valid for a population of
grains with $3 < q < 4$ and $a_{\rm{min}} \ll a_c \ll
a_{\rm{max}}$, with $a_c$ being a function of the real and
imaginary parts of the complex dielectric function and wavelength
(at millimeter wavelengths $a_c \sim 1$ mm). Since, at a fixed
$q$, the function $\beta(a_{\rm{max}})$ decreases when
$a_{\rm{max}} \simgreat 1$ mm (see Figure \ref{fig:dust}), it follows
that the approximated asynthotic value $(q - 3)\beta_{\rm{RJ}}$
gives a lower limit for $\beta$ at a finite $a_{\rm{max}}$.
Adopting $\beta_{\rm{RJ}} = 1.7$ from the small dust grains in the
ISM, for $q\simgreat 3.5$ we derive $\beta \simgreat 0.85$, i.e.
dust grains populations with $q \simgreat 3.5$ cannot explain
$\beta$-values lower than $\approx 0.85$\footnote{Note that
$\beta$-values greater than $\approx 0.85$ can be explained by
both $q \simless 3.5$ and $q \simgreat 3.5$.}.

In Figure \ref{fig:beta_hist} we report the histogram of the $\beta$-values derived by our analysis. The average value in our sample considering both unmapped and resolved disks, $<\beta> = 0.6 \pm 0.06$, is not consistent with dust populations with $q$-values of 3.5, like for the ISM, or greater. Values of $q$ lower than 3.5 have been obtained by, e.g., Tanaka et al.~(\cite{Tan05}) from simulations of grain growth in stratified protoplanetary disks which give vertically integrated size distributions with $q \approx 3$, plus a population of much larger bodies, to which however our observations would not be sensitive.

These considerations clearly suggest that the process of dust grain growth to mm-sizes
has played a fundamental role in most, if not all, of the
Taurus-Auriga protoplanetary disks whose sub-mm/mm SED
has been sufficiently investigated.

From Figure \ref{fig:beta_hist} it is also evident that the two sub-samples made of unmapped and resolved disks separately have different averaged $\beta$ values, with the $\beta$ values being on average lower for the unmapped disks than for the resolved ones ($<\beta> = 0.41\pm0.10$, $0.68 \pm 0.06$ for the unmapped and resolved disks respectively). Since the unmapped sample nearly coincides with the fainter disks with $F_{\rm{1mm}} < 100$ mJy, and since $\beta$ is correlated with the spectral index $\alpha$, this is due to the trend already presented in Figure \ref{fig:flux_alpha} between $F_{\rm{1mm}}$ and $\alpha_{\rm{1-3mm}}$.
This trend agrees with simple expectations for grain growth: as the maximum grain size $a_{\rm{max}}$ increases, both $\beta$ and $\kappa_{\rm{1mm}}$ decrease, so disks with the same dust mass should become fainter as $\beta$ gets smaller. Alternatively, the $F_{\rm{1mm}}$-$\alpha_{\rm{1-3mm}}$ trend could arise if the fainter disks are smaller, and then more optically thick, than the brighter, resolved ones, while their ``true'' $\beta$ would be nearly the same.

\begin{figure}
 \centering
 \resizebox{\hsize}{!}{\includegraphics{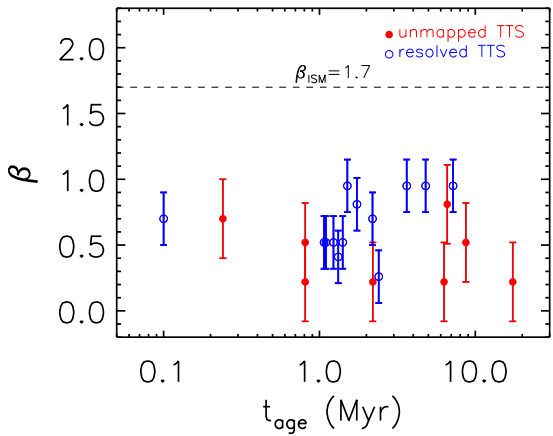}}
 \caption{Beta versus age: relationship between the dust opacity spectral index $\beta$ and the estimated stellar age obtained as in Section \ref{sec:stars}. Open blue points represent the spatially resolved disks, whereas the filled red points are for the unmapped ones. The $\beta$ value for the ISM dust is indicated as a dashed horizontal line.}
 \label{fig:age_beta}
\end{figure}

In Figure \ref{fig:age_beta} we show the $\beta$ vs stellar age plot. No evidence of an evolutionary trend is present over the large range of stellar ages spanned by our sample. As already discussed the determination of the individual ages of PMS stars is very uncertain. Hartmann~(\cite{Har01}) has shown that the large age spread deduced for T Tauri stars in Taurus may just be the result of uncertainties in the measurements towards individual members of a population with a very narrow age spread. Our most solid conclusion, not affected by the age determination uncertainties, is that most (if not all) of the Class II YSO in our sample show evidence of evolved grains in the disk outer regions.

\subsection{Grain growth}
\label{subsec:grain_growth}

These results confirm two statements that have been addressed in the last years. The first one is that the dust coagulation from ISM dust to mm/cm-sized grains is a fast process, that appears to occur \textit{before} a YSO enters in the Class II evolutionary stage, i.e. when a dense envelope still surrounds the protostar+disk system. This is also consistent with the recent results by Kwon et al.~(\cite{Kwo09}) who have found evidence of grain growth in 3 Class 0 YSOs in Perseus and Cepheus through CARMA observations at 1.3 and 2.7 mm.

The second statement is that the mm/cm-sized dust grains appear to stay longer in the outer disk if compared with their expected inward drift timescale due to the interaction with the gas component. In a circumstellar disk the gas experiences a pressure gradient force which cancels a part of the central star's gravity and induces a rotation slower than the Keplerian velocity. Dust grains, whose orbits are hardly affected by gas pressure, are assumed to be rotating with Keplerian velocity, and thus experience a headwind. They lose angular momentum and spiral inward to the central star (Adachi et al.~\cite{Ada76}, Weidenschilling~\cite{Wei77}).
Takeuchi \& Lin~(\cite{Tak05}) proposed as possible solutions to the drift problem that either the grain growth process from mm-sized to cm-sized grains is very long in the outer regions of the disk (timescales greater than 1 Myr, e.g. given by grains' sticking probability as low as 0.1) or that grains have already grown to 10 m or larger and they do not migrate rapidly anymore, and mm-sized particles are continuosly replenished through collisions of these large bodies. However our analysis suggests that, unless the grain size distribution index is very low, i.e. $q < 2.5$, disks with $\beta \simless 0.5$ show evidence of grains as large as $\sim 1$ cm in the outer regions. Also, as already pointed out in Natta et al.~(\cite{Nat04}) and Brauer et al.~(\cite{Bra07}), a noteworthy amount of larger bodies, which would not contribute to the disk emission because of their negligibly small opacity coefficient, could significantly increase the disk masses. Since the typical orbital decay time for a cm-sized grain with a density of 0.1 g cm$^{-3}$ at 50 AU is $\approx 10^{5}$ yr ($10^{4}$ yr for a more compact grain with a 1 g cm$^{-3}$ density), to explain the presence of cm-sized grains in the older disks in our sample (with ages of few Myrs and up to $\sim 10^7$ yr) the reservoir of larger bodies would have to contain at least 100 times more mass than the observed dust mass. The total disk mass would then be comparable to or even exceeding the estimated stellar mass (using a standard gas-to-dust mass ratio of 100, see Section \ref{subsec:disk_mass}).

To solve the drift problem for very low and very high disk masses Brauer et al.~(\cite{Bra07}) proposed a significant reduction of the drag experienced by the dust due to a reduction of the gas-to-dust ratio in the disk from the canonical value of 100, and to possible collective effects of dust gathering in a thin midplane layer in the case of very-low turbulent disks. For example, using a low value for the turbulence parameter $\alpha = 10^{-6}$ and a gas-to-dust ratio of 5, they find that cm-sized dust grains remain in the outer parts of the disk for more than 2 Myr if disk masses $<0.05 \ M_{\star}$ or $>0.2 \ M_{\star}$ are considered. In the range $M_{\rm{disk}} \approx 0.05 - 0.2 \ M_{\star}$ the collective effects of dust are not efficient enough to retain cm-sized grains in the disk outer regions. If the disk is more turbulent, then more gas has to be removed in order to keep the particles in the outer parts for a longer period of time. To justify such small gas-to-dust ratios tha authors suggest the process of photoevaporation as the main driver of gas removal. However, in this case, since photoevaporation is expected to manifest only in the later stages of disk evolution, then we would expect a strong evolutionary trend of $\beta$, which is instead not found by our analysis.

Another possibility is the presence in the disk of local gas pressure maxima that would block the inward drift of the dust grains. As shown by the dynamics calculation of Barge \& Sommeria~(\cite{Bar95}) and Klahr \& Henning~(\cite{Kla97}) this mechanism may provide a solution also to the relative velocity fragmentation barrier that prevents meter-sized bodies to further grow in size in the inner regions of the disk.

\subsection{Dust mass}
\label{subsec:disk_mass}

From the sub-mm/mm SED fitting procedure we derived the quantity
$M_{\rm{dust}} \times
\kappa_{\rm{1 mm}}$. To obtain the dust mass in the disks an
estimate of $\kappa_{\rm{1 mm}}$ is needed. The choice of the dust
opacity at $1$ mm creates the largest uncertainty in the
derivation of (dust) disk masses from millimeter observations.
Estimates of $\kappa_{\rm{1 mm}}$ from different dust models in
the literature span a range of a factor greater than 10. The most
commonly used value in the literature is $\kappa_{\rm{1mm}} \simeq
3$ cm$^2$ per gram of dust (Beckwith et al.~\cite{Bec90}) that, for
simplicity, is supposed to be constant, i.e. independent of the
dust grain properties, for all the disks. As noted in Natta et al.~(\cite{Nat04})
this is not realistic as grain growth affects both the frequency
dependence of the dust opacity and its absolute value (e.g.
Miyake \& Nakagawa~\cite{Miy93}).

In this paper we use the dust model described in Section
\ref{sec:opacity} to get an estimate of $\kappa_{\rm{1 mm}}$ that takes into account the dust physical properties, expecially grain growth. As
shown in the left top panel of Figure \ref{fig:dust}, at a fixed value
of the grain size distribution index $q$, the value of
$\kappa_{\rm{1 mm}}$ can be inferred from $a_{\rm{max}}$ which in
turn can be derived from $\beta$ through the relation plotted in
the left bottom panel of the same figure. The relation between $\kappa_{\rm{1 mm}}$ and $\beta$ is shown in the right panel. Therefore, from an estimate
of $\beta$ and $M_{\rm{dust}} \times \kappa_{\rm{1mm}}$ obtained by the sub-mm/mm
SED fitting procedure we can get an estimate for $\kappa_{\rm{1
mm}}$ and then the dust mass $M_{\rm{dust}}$ for our dust
model, after fixing the $q$ parameter. In Table
\ref{tab:fits_results} we report the values of $M_{\rm{dust}}$ as
obtained with this method for $q = 2.5$, $3.0$, $3.5$. It is important to bear in mind that, as for the estimate of the maximum grain size, also for $\kappa_{\rm{1 mm}}$, and thus for the dust mass, the estimates depend on the specific dust model that one adopts (see discussion in Natta et al.~\cite{Nat04}). However, following Isella et al.~(\cite{Ise09}), we take into account the variation of the dust opacity with $\beta$ using a well defined dust model to infer dust masses, and this approach is more accurate than considering a fixed $\kappa_{\rm{1 mm}}$ as often done in previous studies.

In general, for our specific dust model, when $\beta < 1$ then $M_{\rm{dust}}^{q=2.5} < M_{\rm{dust}}^{q=3} <
M_{\rm{dust}}^{q=3.5}$, the amount of variation of
$M_{\rm{dust}}^{q}$ for the different $q$ values being dependent
on $\beta$. In particular, when $\beta \simeq 0.8-1$
$M_{\rm{dust}}^{q=2.5}$ and $M_{\rm{dust}}^{q=3}$ are nearly
coincident and $M_{\rm{dust}}^{q=3.5}$ is greater than
$M_{\rm{dust}}^{q=2.5}$ by factors of $\sim 2-4$. For $\beta
\simless 0.7$ a grain size distribution with $q = 3.5$ cannot
explain the observational data and the discrepancy between
$M_{\rm{dust}}^{q=2.5}$ and $M_{\rm{dust}}^{q=3}$ increases up to
a factor of $\sim 15$ when $\beta \simeq 0.2$. This is due to the increase of the distance between the $\kappa_{\rm{1mm}}$($\beta$) curves at fixed $q$-values, when $\beta$ decreases, as shown in the right panel of Figure \ref{fig:dust}

As already mentioned in Section \ref{subsec:spectral_slopes}, from the value of $\beta$ it is
impossible to disentangle $a_{\rm{max}}$ and $q$.
Nevertheless, only the disks showing $\beta \geq 0.8$ can be
explained by a grain size distribution index $q = 3.5$. For this
reason, for the rest of our analysis we will consider separately
only the cases in which the dust grain size distribution is
characterized by $q = 2.5$ or $3$. The value of $q$ in
protoplanetary disks is unknown and it is not observationally
constrainable, and so it is in principle possible that it can vary
substantially from one disk to the other. For our discussion we
will use the assumption that $q$ is the same for all the disks.

Once the dust mass is estimated, this is commonly converted into an
estimate of the total disk mass (dust$+$gas) using the standard
ISM gas-to-dust ratio of 100. However, if the dust and gas
components have different evolutionary timescales, this ratio is a
function of time and the standard ISM value would be inappropriate
for estimating the disk mass at the present time. Physical
processes that are likely to alter this ratio during the evolution
of a protoplanetary disk are gas photo-evaporation, which would
lead to a decrease of the gas-to-dust ratio (unless a significant
amount of dust is dragged away by the gas flow), and dust
inward migration, which instead would increase it.
Natta et al.~(\cite{Nat04}) suggested that the standard ISM ratio provides a correct estimate
of the ``original'' disk mass, when the disk composition reflected
that of the parent cloud. However, this is true only if a negligible amount of dust has migrated with respect to the gas and accreted onto the central star, otherwise the ISM ratio would provide only a lower limit to the original disk mass. Because of our ignorance on the gas-to-dust ratio evolution in this paper we will not convert the dust mass into a total disk mass.
Also, it is important to note that since millimeter observations are insensitive to the emission of bodies much larger than 1 cm, including planetesimals and planets, the dust mass we are referring to in this paper does not include the contribution from these larger objects.  The dust mass is intended to be the mass in ``small grains'' only, and is a lower limit to the total mass in dust.

\begin{figure*}[hptb!]
 \centering
 \begin{tabular}{cc}
 \includegraphics[scale=0.45]{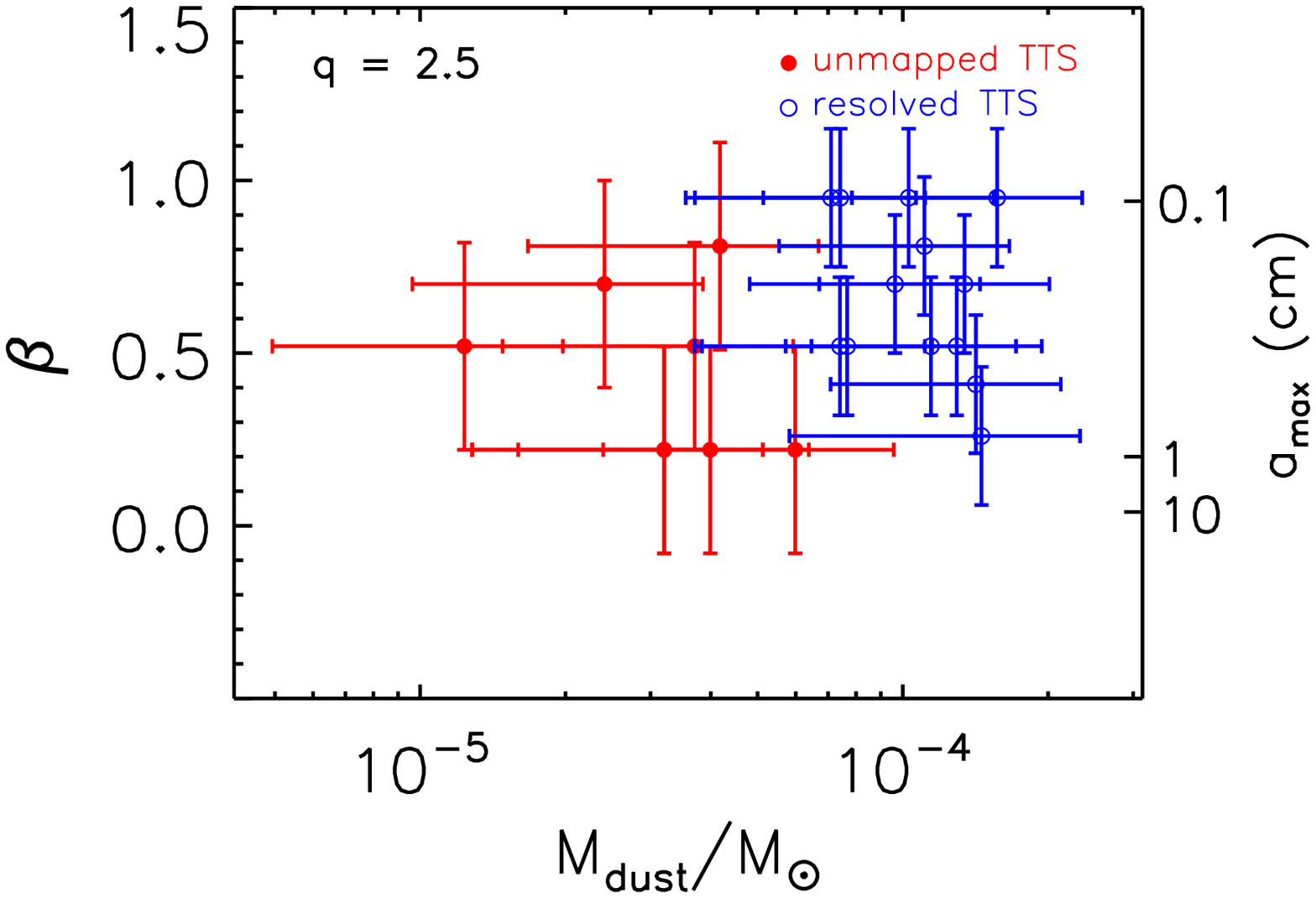} &
 \includegraphics[scale=0.45]{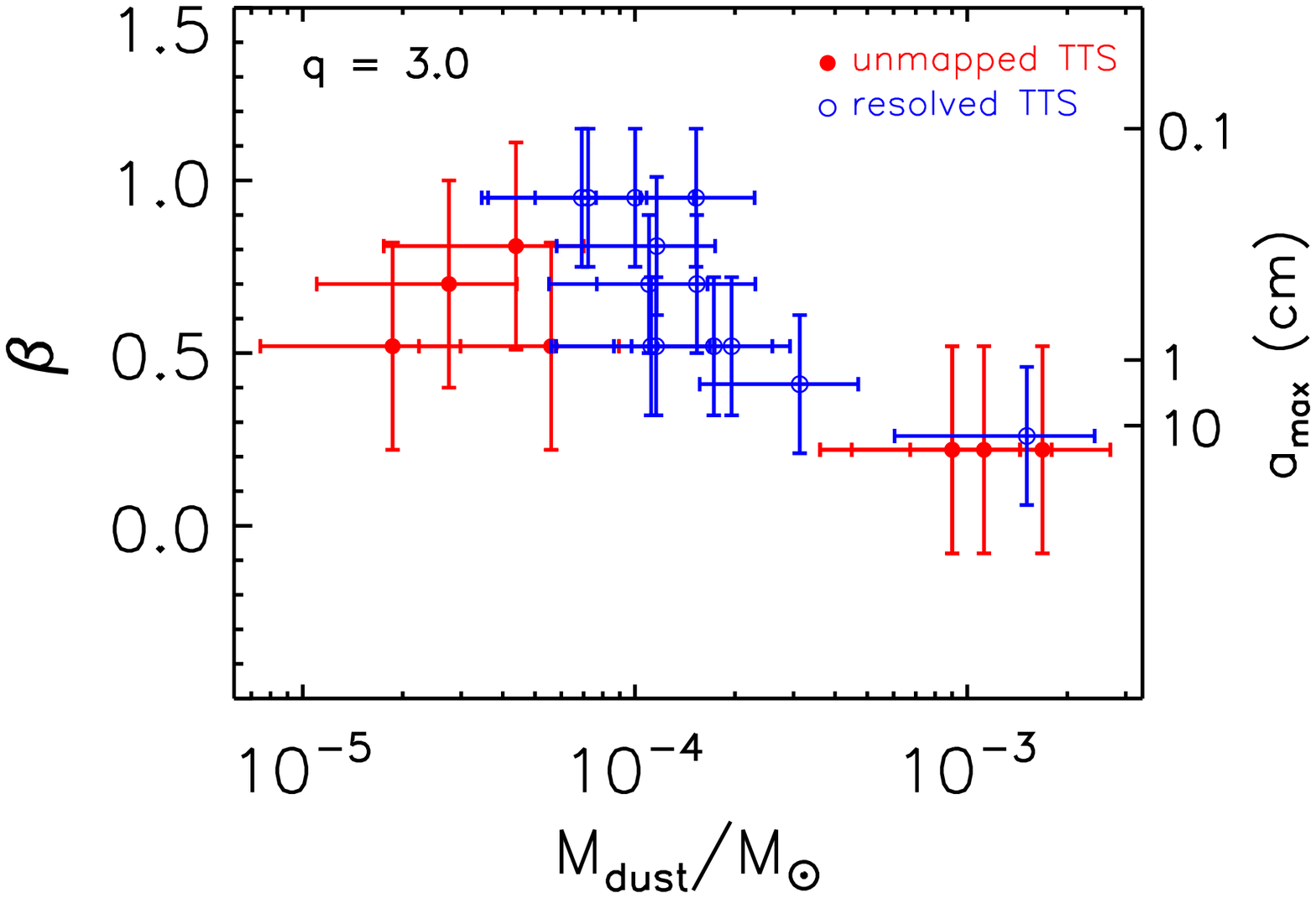} \\

 \end{tabular}
\caption{Dust opacity spectral index $\beta$ and maximum grain
size a$_{\rm{max}}$ plotted against the disk mass in dust for the
sources in our sample. In the left and right panels the dust mass
has been obtained adopting a value of $2.5$ and $3$ respectively
for the power index $q$ of the dust grains size distribution. Open blue
points represent the spatially resolved disks, whereas the filled red
points are for the unmapped ones. The errorbars in the dust mass
estimates do not take into account our uncertainty on the real
value of $q$.}
 \label{fig:disk_mass_beta}
\end{figure*}

Figure \ref{fig:disk_mass_beta} shows the $\beta$ vs $M_{\rm{dust}}$ plots for $q$ equal to 2.5 and 3. The plot with $q = 2.5$ does not show any significant relation.  In the $q = 3$ case an anticorrelation between these two quantities may be present, with the more massive dust disks being associated to the ones showing today the largest grains in their outer parts, with $a_{\rm{max}}$ up to $\sim 10$ cm. This possible anticorrelation is clearly an effect of the choice of a physical model for the dust that takes into account the variation of the dust opacity normalization factor with the size-distribution of the dust grains (Section \ref{sec:opacity}). If we had made the unrealistic assumption of a constant value for the dust opacity at, say, 1 mm, we would have found an opposite result, i.e. disks with \textit{less} mass in dust would have shown on average larger grains, as can be derived from Figure \ref{fig:flux_alpha}, since $\alpha$ is approximately proportional to $\beta$ and $F_{\rm{1 mm}}$ would be approximately proportional to the dust mass, in the constant $\kappa_{\rm{1 mm}}$ case. More realistically, lower values of $\beta$ imply low absolute values of the dust millimeter opacity (see right panel of Figure \ref{fig:dust}), so that, for a fixed observed flux and disk temperature, the corresponding dust mass must be larger. The reason why the plot with $q = 2.5$ does not show signs of anticorrelation is that at a fixed $\beta$ a lower $q$ value is associated to a smaller dust grain maximum size.  The $\beta$ range spanned by the disks in our sample translates into a range in $\kappa_{\rm{1mm}}$ that is narrower for $q = 2.5$ than it is for $q = 3$ (see right panel of Figure \ref{fig:dust}), thus leading to smaller correction factors with respect to the constant $\kappa_{\rm{1mm}}$ case. Note that among the 5 disks showing the smallest $\beta$ ($\approx 0.2-0.3$) and the highest dust masses (about $10^{-3} \ M_{\odot}$ for $q = 3$) only one (AA Tau) has been mapped and spatially resolved. High angular resolution and high sensitivity imaging of these disks together with other ones are needed to test the robustness of the $\beta-M_{\rm{disk,init}}$ anticorrelation in the $q = 3$ case.

\begin{figure*}[htb!]
 \centering
 \begin{tabular}{cc}
 \includegraphics[scale=0.45]{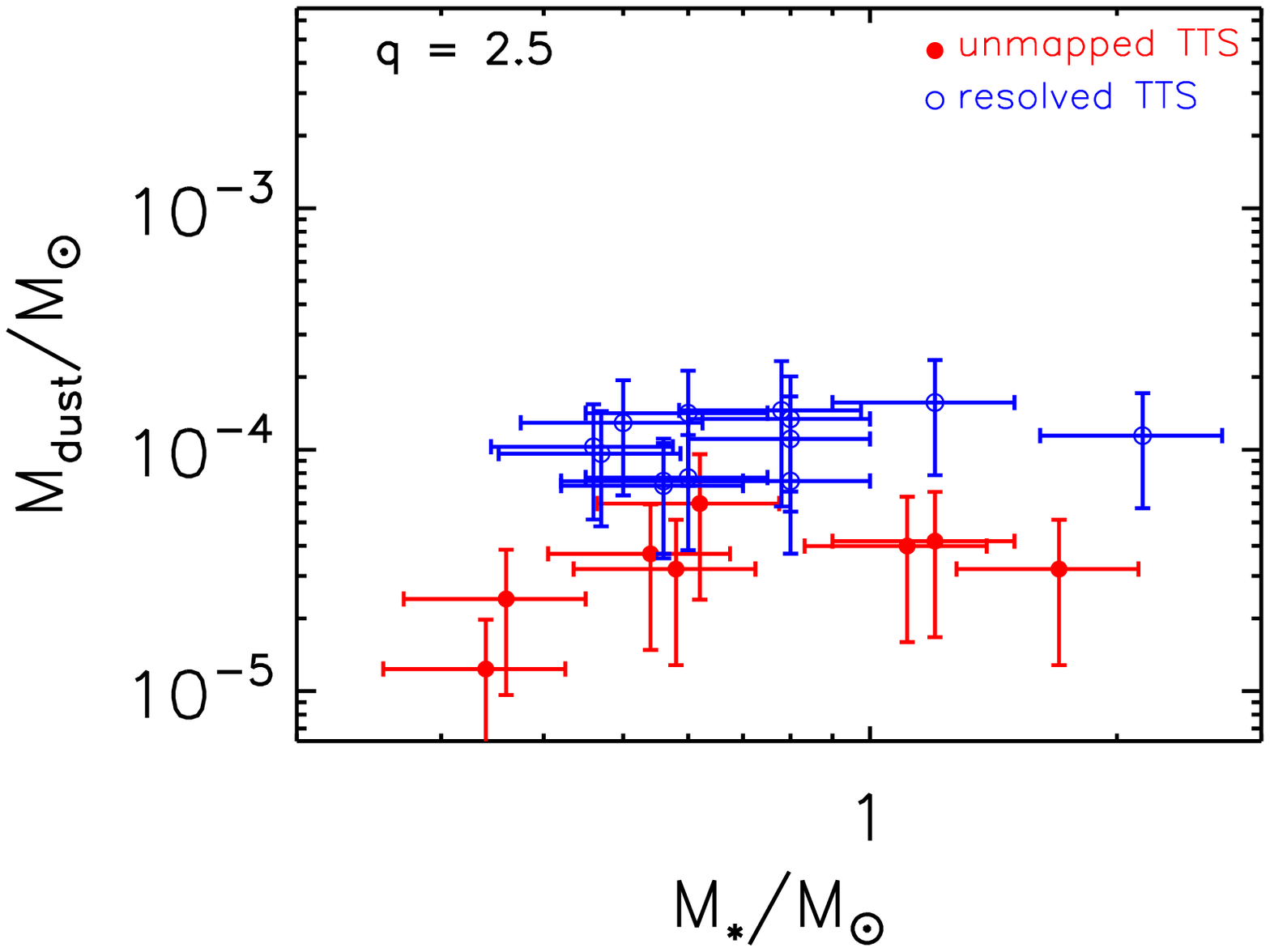} &
 \includegraphics[scale=0.45]{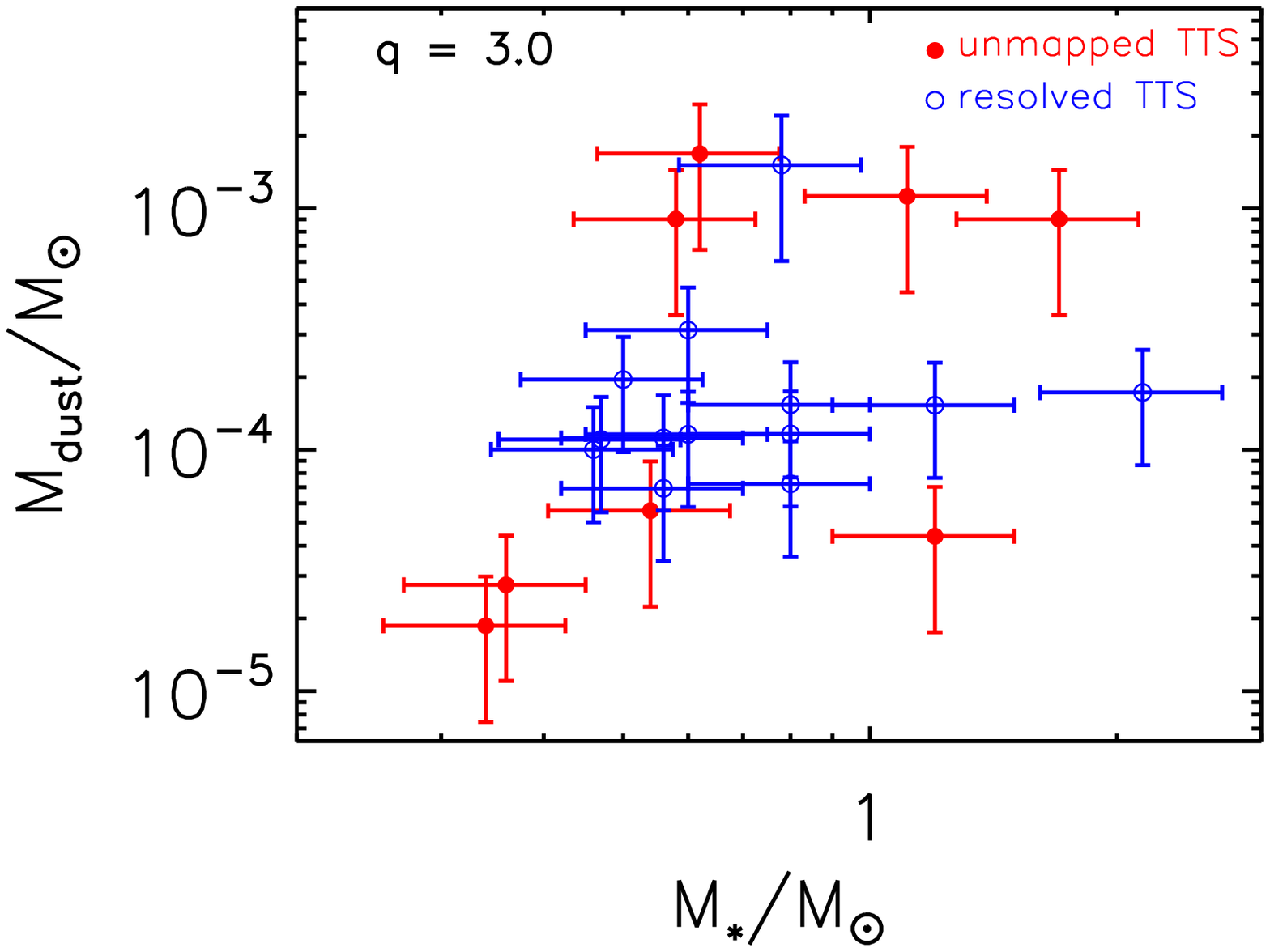} \\

 \end{tabular}
\caption{Disk mass in dust plotted against the stellar mass for the sources in our sample. In the left and right panels the dust mass has been obtained adopting a value of $2.5$ and $3$ respectively for the power index $q$ of the dust grains size distribution. The stellar mass has been estimated as described in Section \ref{sec:stars}, and the errorbars take into account the range of values given by different PMS stars evolutionary models. Open blue points represent the spatially resolved disks, whereas the filled red points are for the unmapped ones. The errorbars in the dust mass estimates do not take into account our uncertainty on the real value of $q$.}

 \label{fig:stellar_mass_disk_mass}
\end{figure*}

\begin{figure*}[htb!]
 \centering
 \begin{tabular}{cc}
 \includegraphics[scale=0.45]{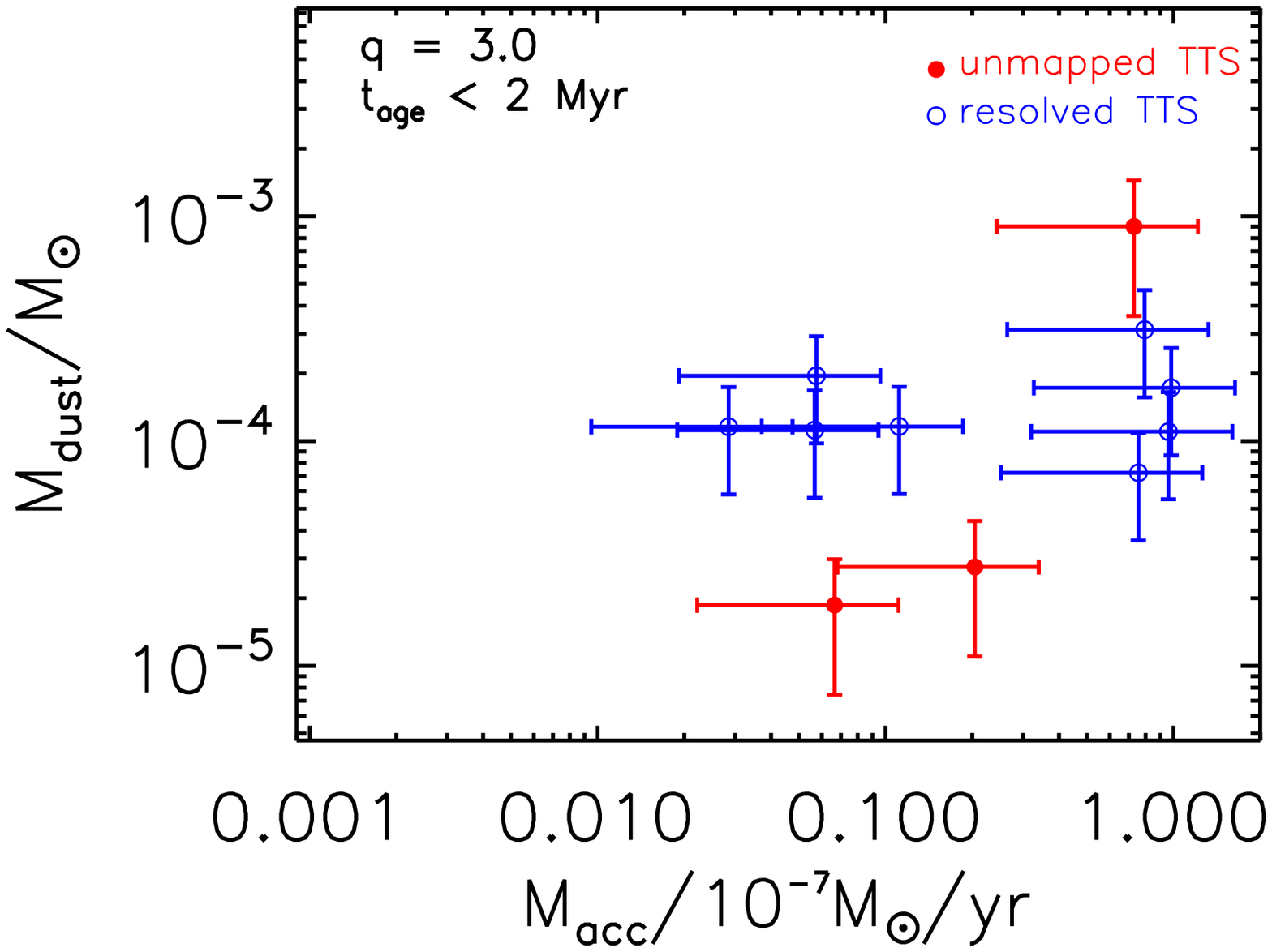} &
 \includegraphics[scale=0.45]{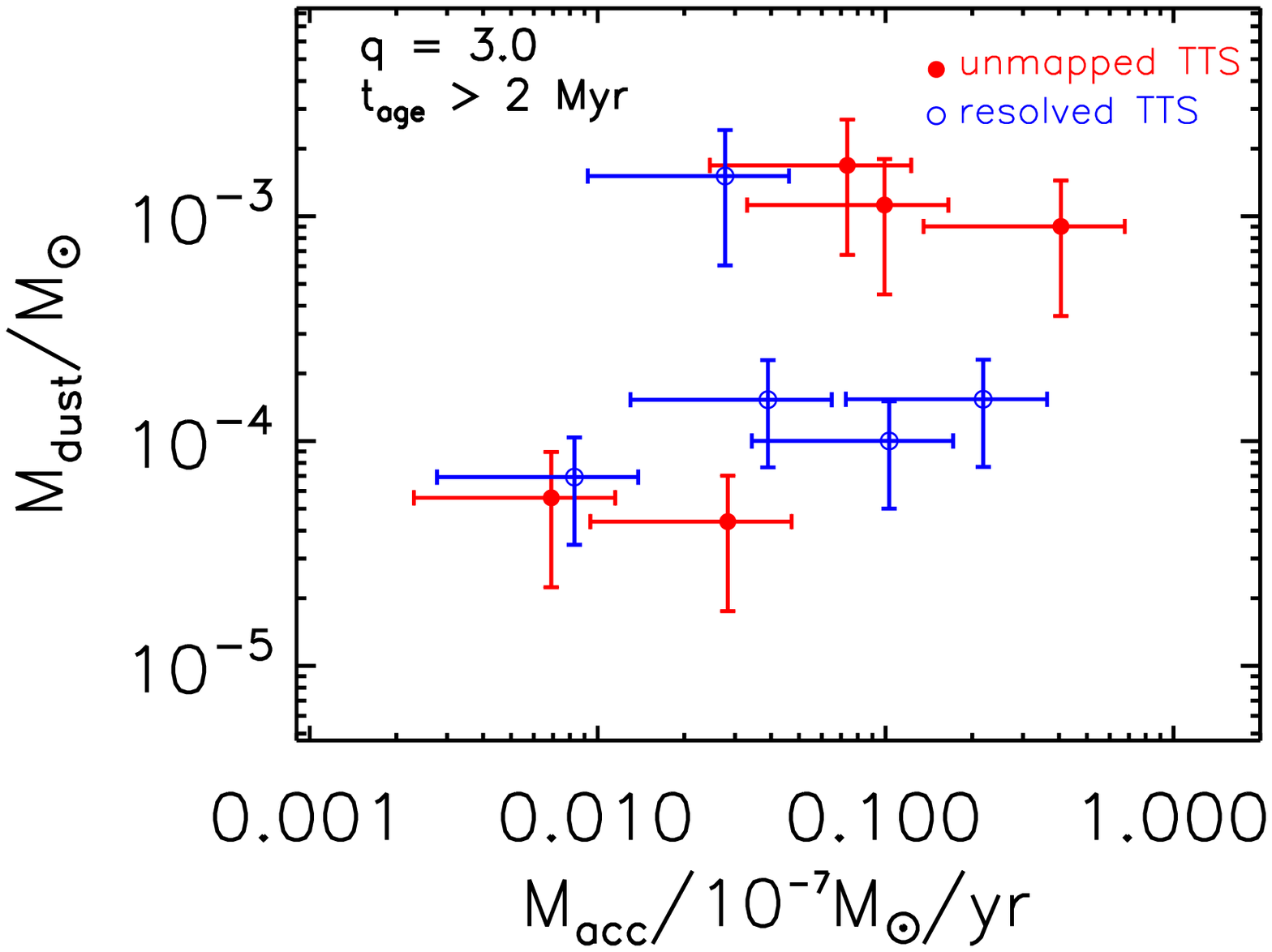} \\

 \end{tabular}
\caption{Disk mass in dust plotted against the mass accretion rate for the sources in our sample. In the left and right panels we included the objects with estimated ages of less and more than 2 Myr respectively. The dust mass has been obtained adopting a value of $3$ for the power index $q$ of the dust grains size distribution. The stellar mass accretion rate has been estimated following the method described in Section \ref{sec:mdot}. Open blue points represent the spatially resolved disks, whereas the filled red points are for the unmapped ones. The errorbars in the mass accretion rate take into account the typical intrinsic time-variability of the accretion rates. The errorbars in the dust mass estimates do not include our uncertainty on the actual value of $q$.}

 \label{fig:mass_acc_disk_mass_3}
\end{figure*}

Figure \ref{fig:stellar_mass_disk_mass} and \ref{fig:mass_acc_disk_mass_3} plot the disk mass in dust against the estimated mass of the central star (dust mass obtained with $q$-values of 2.5 and 3) and mass accretion rate (dust mass with $q = 3$ and stellar ages smaller and greater than 2 Myr) respectively. In the $0.3-2 \ M_{\odot}$ range in stellar mass the dust mass does not show any trend (for $q = 2.5$: $3\cdot 10^{-5} < M_{\rm{dust}} < 2\cdot 10^{-4} M_{\odot}$; for $q = 2.5$: $4\cdot 10^{-5} < M_{\rm{disk,init}} < 2\cdot 10^{-3} M_{\odot}$). The two YSOs with $M_{\star} \simless 0.3 \ M_{\odot}$ have lower dust masses than all the other YSOs for both $q =$ 2.5 and 3. However, the range in stellar mass spanned by our sample is rather limited, and particularly biased toward YSOs with late-K and early-M spectral types. From the upper limits obtained at 1.3 mm by Schaefer et al.~(\cite{Sch09}) for 14 T-Tauri stars with spectral types later than $\approx$ M2  ($F_{\rm{1.3mm}} \simless 5-20$ mJy), and from the detection of very low disk masses around sub-stellar objects in Taurus-Auriga (Klein et al.~\cite{Kle03}, Scholz et al.~\cite{Sch06}), it is likely that a correlation between the disk mass in dust and the mass of the central star exists. High sensitivity millimeter observations are needed to constrain the disk masses in dust around very low mass PMS stars. In Figure \ref{fig:mass_acc_disk_mass_3} a trend of higher mass accretion rates for the more massive disks in dust may be present in the data but still the scatter is very large (a similar result is obtained for the $q = 2.5$ case).
Other relations between the dust properties ($\beta$, dust mass) and the stellar ones (e.g. stellar luminosity, effective temperature) have been investigated but no significant correlations have been found.

Finally, we have compared the values of $\beta$ obtained in this paper with the ones derived in Natta et al.~(\cite{Nat04}) for a sample of 6 intermediate mass Herbig Ae stars and 3 lower mass T Tauri stars. From the two-sided Kolmogorov-Smirnov test, the probability that the $\beta$ values from the two samples come from the same distribution is about $88\%$, indicating that the disks in the two samples have similar dust properties in terms of grain growth (in the Natta et al.~\cite{Nat04} sample only one object, the Herbig Ae star HD150193, has a relatively high $\beta = 1.6 \pm 0.2$, compatible with non-evolved ISM-like dust grains).

\section{Summary}
\label{sec:summary}

We have analysed the sub-mm/mm SED out to 3mm of a sample of 21 Class II T-Tauri stars in the Taurus-Auriga star forming region. This sample comprises about the $60\%$ of the ``isolated'' class II YSOs with an estimated stellar mass greater than $\approx 0.4 \ M_{\odot}$. It also comprises approximately $1/3$ of the isolated class II wich have been observed at 0.85mm to have a flux less than 100 mJy, for which we provide first 3mm detections thanks to sensitive PdBI observations. Our main findings are summarized below:
\begin{enumerate}
\item For all the sources in our sample the millimeter opacity spectral index $\beta$ is significantly lower than the value obtained for the dust in the ISM. For the 13 spatially resolved disks this is a clear evidence of the presence in the outer disk regions of dust grains as large as at least 1 mm. For the 8 unmapped sources, assuming disk outer radii comparable to the resolved sources low $\beta$ values are found as well, suggesting that even for these fainter sources 1 mm-sized grains are in the disk outer regions. This confirms, over a larger sample and less massive disks, the results obtained by past observations on T-Tauri stars (e.g. Rodmann et al.~\cite{Rod06} and references therein).
\item No significant evidence of an evolutionary trend for the millimeter opacity spectral index $\beta$ has been found. This indicates that the dust grain growth to $\sim 1$ mm-sizes is a very fast process in a protoplanetary disk, that appears to occur before a YSO enters in the Class II evolutionary stage. Also, the amount of these large grains in the disk outer regions does not appear to decline throughout all the Class II evolutionary stage.
\item Only for 9 sources in our sample a power law index for the grain size distribution of the ISM, $q_{\rm{ISM}} \approx 3.5$, is consistent with the sub/mm-mm SED. Instead, most of the disks show evidence of a smaller $q-$value.
\item We have not found any significant correlation between the dust properties, namely grain growth and dust mass in the disk outer regions, and the properties of the central star, including the mass accretion rate onto the stellar surface. However, our sample contains mostly YSOs with a narrow range in spectral types, namely between late-K and early-M, and it is possible that a correlation between the disk mass in dust and the stellar mass will become apparent when considering also disks around late-M spectral types.
\item The 6 faintest sources in our sample, with $F_{\rm{1mm}} < 50$ mJy, show a spectral index between 1 and 3 mm that is on average lower than the spectral index of the brighter sources. This may indicate either that for these fainter, yet unmapped disks the emission from the optically thick inner disk is much more significant than for the brighter sources or that the dust grains in the outer regions of these fainter disks are even larger than in the brighter YSOs, possibly up to 10 cm if $q = 3$. The latter hypothesis would imply on average a lower absolute value for the millimeter opacity coefficient for these fainter disks. Taking into account this effect through a physical model for the dust properties, we found, in the case of a value for the power index of the grain size distribution $q = 3$, an anticorrelation between $\beta$ and the disk dust mass, that would indicate that larger dust grains are found in the outer parts of the more massive disks (in dust).
\end{enumerate}

In the next future, with the advent of the next sub-mm/mm facilities (both new interferometers like ALMA and technical improvements in the existent arrays, like for example the E-VLA) it will be possible to extend the investigation of the dust properties to fainter circumstellar disks around the lowest mass PMS stars and brown dwarfs. Furthermore, thanks to the great enhancements in both high angular resolution and sensitivity that these facilities will provide, it will be possible to determine their spatial extension.

\begin{acknowledgements}
We thank the anonymous referee for his/her comments that greatly improved the manuscript. L.R. acknowledges the PhD fellowship of the International Max Planck Research School.
\end{acknowledgements}



\end{document}